\documentclass[]{pasj02} 

\usepackage{natbib}
\usepackage[switch,mathlines]{lineno} 
\usepackage{comment}
\usepackage[T1]{fontenc}
\usepackage[]{rotating}
\usepackage{graphicx}
\usepackage{caption} 
\usepackage{subcaption}
\usepackage{tablefootnote}
\usepackage[T1]{fontenc} 
\usepackage[utf8]{inputenc} 
\usepackage{etoolbox}
\usepackage{flushend}
\usepackage{balance}
\usepackage{multicol}
\usepackage{newunicodechar}
\usepackage{graphicx}
\usepackage{subcaption}
\usepackage{pdflscape}
\usepackage{caption}

\newunicodechar{ʻ}{\textquoteleft}

\jyear{2026}
\Received{}

\begin{document} 

\title{Discovery of four Narrow-line Type-1 Quasars at z = 2.54 -- 6.06  with Subaru \textquoteleft\={O}nohiʻula PFS-SSP}

\author{
 Ayumi \textsc{Takahashi},\altaffilmark{1}\altemailmark\orcid{0000-0003-3769-6630}\email{ayumi@kanagawa-u.ac.jp}
 Yoshiki \textsc{Matsuoka},\altaffilmark{2}\orcid{0000-0001-5063-0340}
 Yoshiki \textsc{Toba},\altaffilmark{2,3,4}\orcid{0000-0002-3531-7863}
 Tohru \textsc{Nagao},\altaffilmark{2,5}\orcid{0000-0002-7402-5441}
 Toshihiro \textsc{Kawaguchi},\altaffilmark{6}\orcid{0000-0002-3866-9645}
 Hiroaki \textsc{Sameshima},\altaffilmark{7}\orcid{0000-0001-6401-723X}
 Alberto \textsc{Rodr\'{\i}guez-Ardila},\altaffilmark{8}\orcid{0000-0002-7608-6109}
 Andy D. \textsc{goulding},\altaffilmark{9}\orcid{0000-0003-2871-127X}
 Masayuki \textsc{Akiyama},\altaffilmark{10}\orcid{0000-0002-2651-1701}
 Denimara \textsc{Dias dos Santos},\altaffilmark{11}\orcid{0000-0003-4153-4829}
 Atsushi \textsc{Hoshi},\altaffilmark{12}\orcid{0009-0003-7134-0539}
 Kohei \textsc{Ichikawa},\altaffilmark{10}\orcid{0000-0002-4377-903X}
 Masatoshi \textsc{Imanishi},\altaffilmark{13,14}\orcid{0000-0001-6186-8792}
 Kazushi \textsc{Iwasawa},\altaffilmark{15,16}\orcid{0000-0002-4923-3281}
 Mitsuru \textsc{Kokubo},\altaffilmark{13}\orcid{0000-0001-6402-1415}
 Huynh Anh N. \textsc{Le},\altaffilmark{17}\orcid{0000-0003-1270-9802}
 Khee-Gan \textsc{Lee},\altaffilmark{18,19}\orcid{0000-0001-9299-5719}
 Masafusa \textsc{Onoue},\altaffilmark{20}\orcid{0000-0003-2984-6803}
 Masayuki \textsc{Tanaka},\altaffilmark{13,14}\orcid{0000-0002-5011-5178}
 Patrice \textsc{Theul\'e},\altaffilmark{21}
 Hao \textsc{Zhang},\altaffilmark{21}
Jialai \textsc{Wang},\altaffilmark{17,18,19,22}\orcid{0009-0001-9170-3363}
Xinfeng \textsc{Xu},\altaffilmark{23}\orcid{0000-0002-9217-7051}
Yongquan \textsc{Xue}\altaffilmark{17}
}

\altaffiltext{1}{Faculty of Science, Kanagawa University, 3-27-1 Rokukakubashi, Kanagawa-ku, Yokohama, Kanagawa 221-8686, Japan}
\altaffiltext{2}{Research Center for Space and Cosmic Evolution, Ehime University, 2-5 Bunkyo-cho, Matsuyama, Ehime 790-8577, Japan}
\altaffiltext{3}{Department of Physical Sciences, Ritsumeikan University,
1-1-1 Noji-higashi, Kusatsu, Shiga 525-8577, Japan}
\altaffiltext{4}{Academia Sinica Institute of Astronomy and Astrophysics,
11F of Astronomy-Mathematics Building, AS/NTU, No.1, Section 4,
Roosevelt Road, Taipei 10617, Taiwan}
\altaffiltext{5}{Amanogawa Galaxy Astronomy Research Center, Kagoshima University, 1-21-35 Korimoto, Kagoshima 890-0065, Japan}
\altaffiltext{6}{Graduate School of Science and Engineering, University of Toyama, Toyama 930-8555, Japan}
\altaffiltext{7}{Institute of Astronomy, Graduate School of Science, The University of Tokyo, 2-21-1 Osawa, Mitaka, Tokyo 181-0015, Japan}
\altaffiltext{8}{Observatório Nacional, Rua General José Cristino 77, CEP 20921-400, Rio de Janeiro, RJ, Brazil}
\altaffiltext{9}{Department of Astrophysical Sciences, Princeton University, Princeton, NJ 08544, USA}
\altaffiltext{10}{Astronomical Institute, Tohoku University, Aoba, Sendai, 980-8578, Japan}
\altaffiltext{11}{Divis\~{a}o de Astrof\'{\i}sica, 
Instituto Nacional de Pesquisas Espaciais (INPE), 
Avenida dos Astronautas 1758, 
S\~{a}o Jos\'e dos Campos, 
12227-010, SP, Brazil}
\altaffiltext{12}{Department of Physics, Graduate School of Science, Nagoya University,
Furocho, Chikusa, Nagoya 464-8602, Japan}
\altaffiltext{13}{National Astronomical Observatory of Japan, National Institutes of Natural Sciences, 2-21-1 Osawa, Mitaka, Tokyo 181-8588, Japan}
\altaffiltext{14}{ Department of Astronomical Science, Graduate University for Advanced Studies (SOKENDAI), Mitaka, Tokyo 181-8588, Japan}
\altaffiltext{15}{Institut de Ci\`encies del Cosmos (ICCUB), Universitat de Barcelona (IEEC-UB), Mart\'{\i} i Franqu\`es, 1, 08028 Barcelona, Spain}
\altaffiltext{16}{ICREA, Pg Llu\protect\'{\i}s Companys 23, 08010 Barcelona, Spain}

\altaffiltext{17}{Department of Astronomy, University of Science and Technology of China, Hefei 230026, People’s Republic of China}
\altaffiltext{18}{Kavli IPMU (WPI), UTIAS, The University of Tokyo, Kashiwa, Chiba 277-8583, Japan}
\altaffiltext{19}{Center for Data-Driven Discovery, Kavli IPMU (WPI), UTIAS, The University of Tokyo, Kashiwa, Chiba 277-8583, Japan}
\altaffiltext{20}{Waseda Institute for Advanced Study (WIAS), Waseda University, Shinjuku, Tokyo 169-0051, Japan}
\altaffiltext{21}{Aix Marseille Univ, CNRS, CNES, LAM, Marseille, France}
\altaffiltext{22}{School of Astronomy and Space Science, University of Science and Technology of China, Hefei 230026, China}
\altaffiltext{23}{Department of Physics and Astronomy, Northwestern University, 2145 Sheridan Road, Evanston, IL, 60208, USA}




\maketitle
    
\begin{abstract}
    We report the discovery of four narrow-line type-1 quasars (NLQ1s) from the Subaru \textquoteleft\={O}nohiʻula Prime Focus Spectrograph Subaru Strategic Program (PFS-SSP). The obtained spectral properties of the NLQ1s are reminiscent of narrow-line Seyfert 1 (NLS1) galaxies known in the local universe. NLS1 galaxies are luminously powered by their central black holes and exhibit peculiar spectral shapes.
    Since this class has been classified using H$\beta$ and [O\,\textsc{iii}] lines, statistical studies of NLS1s at $z > 2$ remain largely unexplored due to the observational limitations of ground-based telescopes. To extend a comprehensive understanding of this population to the more distant universe, we searched for broad-line AGNs (quasars) whose rest-UV spectral shapes resemble those of local NLS1 galaxies, utilizing the Ly$\alpha$ line profile. We identified four NLQ1s, including one candidate, at $z=2.54$ -- $6.06$. The obtained spectra cover 3800\,\AA\ to 12600\,\AA\, in the observed frame, and show a prominent Ly$\alpha$ emission line with a very low continuum level. The measured full width at half maximum (FWHM) of Ly$\alpha$ ranges from $1410 \ \mathrm{km\,s^{-1}}$ to $1510 \ \mathrm{km\,s^{-1}}$ with a mean of $\sim 1430 \,\mathrm{km\,s^{-1}}$. Although we used only line width information for the NLQ1 selection, the resulting equivalent widths (EW) of Ly$\alpha$ are higher (the mean $\sim 190$\,\AA \ in the rest frame) than the typical values of quasars. Two quasars have exceptionally large Ly$\alpha$ luminosity, accounting for  $\sim 3$ -- $4$\% of the bolometric luminosity. We measured black hole masses and Eddington ratios for three objects whose C\,\textsc{iv} line is available. The resulting black hole masses are $\sim 10^{6.9}$ -- $10^{7.7}\, M_{\odot}$ with high Eddington ratios ($> 0.1$), consistent with those of NLS1 galaxies reported in the low-$z$ universe. N\,\textsc{v} and C\,\textsc{iv} broad absorption lines are detected in one object. This quasar can be classified as a mini-BAL quasar, which further supports the idea that this NLQ1 is indeed a luminous version of an NLS1 galaxy. This paper presents the first statistical search of NLQ1s at high redshifts. With the discovery of these rare objects out to $z \sim 6$, we can accelerate our understanding of the early growth of supermassive black holes.
\end{abstract}


\section{Introduction}

NLS1s are a subclass of active galactic nuclei (AGNs) with small line widths and relatively weak continuum emissions compared to those of typical AGNs. NLS1s have classically been classified using the line width of the broad component of Balmer lines (H$\beta < 2000 \rm{\ km\ s^{-1}}$ in FWHM) with a flux ratio of [O III]/H$\beta < 3$ \citep{Osterbrock1985,Goodrich1989}. They also show strong Fe II emission, soft X-ray spectra, rapid flux variability in X-rays \citep[e.g.,][]{Boller1996,Leighly1999}, and strong outflows \citep[e.g,][]{Boroson2002,Komossa2008, Grupe2010, Xu2012,Paliya2024}.
NLS1s are widely known to have small BH masses ($M_{\rm{BH}}\sim 10^{6-8} M_{\odot}$) with higher Eddington ratios \citep[e.g.,][]{grupe2004,Zhou2006,Xu2012,Rakshit2021} or super Eddington ratio \citep{kawaguchi2003}. A number of models have been proposed to explain the multi-wavelength properties of NLS1 galaxies. These include, for example, the low mass BH-high Eddington ratio relation, connection with outflowing winds and gas density \citep[e.g.,][]{Lawrence1998,xu2007}, as well as metallicity \citep[e.g.,][]{Shemmet2002, Nagao2002} and ionized absorption \citep[e.g.,][]{done2007}. 
It is also worth noting that some studies argue the possibility of viewing angle \citep[e.g.,][]{Decarli2008}. Meanwhile, some observational studies have shown that NLS1 host galaxies tend to have a higher fraction of bar structures \citep{crenshaw2003, ohta2007}.

It remains an open and important question whether NLS1 galaxies represent a distinctive evolutionary phase of SMBH growth, or whether their observed properties are simply a consequence of geometrical projection effects. It is possible, however, that both aspects are simultaneously at work, further complicating the picture.
\citet{Paliya2024} released a large NLS1 catalog from the Sloan Digital Sky Survey Data Release 17 \citep[SDSS DR17;][]{Abdurrou2022}. A total of 22656 NLS1s are compiled in the catalog. Their measured properties confirmed that NLS1 galaxies are powered by highly accreting low-mass BHs.

Fascinating finding of quasars with BH masses of $>10^{8}\ \rm{M_{\odot}}$ at $z \gtrsim 6$ poses a challenge to our understanding of how BHs grew from their seeds within a short timescale \citep{mortlock2011,wu2015,yang2021,wang2021}. However, such mature quasars have already lost the memory of their early BH growth. JWST detected many of the less massive BHs at $z >$ 4, and those BHs show 10-100 times larger masses than expected from the local $M_{\rm BH}-M_*$ relation \citep{maiolino2024,  kocevski2023, ubler2023, harikane2023,furtak2024, pacucci2023, yue2024, stone2024}. The objects residing in the BH-undermassive region of the high-$z$ $M_{\rm BH}-M_*$ plane may be the progenitors of the NLS1 galaxies discovered in the local universe. What physical process drives black holes to evolve onto the local $M_{\rm BH}-M_*$ relation? In light of the above observational evidence, mass accretion at high Eddington ratios is considered to be one of the key processes for BH growth at high redshift. Indeed, various theoretical studies predict that super-Eddington accretion is a plausible mechanism for the origin of supermassive black holes at high redshifts \citep{kawaguchi2004, volonteri2005, wyithe2012,madau2014, alexander2014,shirakata2020}.
Identifying low-mass and high-Eddington-accretion SMBHs, and diagnosing the properties of their host galaxies, is crucial for mapping out the formation and evolution of both black holes and galaxies. 
Therefore, searching for high-$z$ NLS1s, not typical quasars, i.e., narrow-line Type-1 quasars (NLQ1s; bright analogous to NLS1s, \citealt{komossa2006, ji2015}) at high-$z$, is needed.
However, classically, NLS1s have been classified only up to $z \sim 1$ due to the limited wavelength coverage of ground-based telescopes, which prevents the observation of the H$\beta$ and [O\,\textsc{iii}] lines at redshifted to near-infrared wavelength. \citet{Rakshit2021} presented a catalog of NLS1 candidates beyond $z=0.8$ built from SDSS DR14 quasars with the criteria of the Mg\,\textsc{ii} line width, based on the strong correlation between the Mg\,\textsc{ii} and H\,$\beta$ emission line widths reported in \citet{rakshit2020}. They identified as many as 2684 NLS1 candidates at $z=0.8$ -- $2.5$. They also selected 3109 low-$z$ NLS1s ($z < 0.8$) in the SDSS DR14 quasar catalog with the classical definition of FWHM (H$\beta$) < 2000 $\mathrm{km\ s^{-1}}$ and [O\,\textsc{iii}]/H$\beta$ < 3. However, the latter criterion could only reduce 22 sources; this indicates that [O\,\textsc{iii}]/H$\beta$ is not a major factor in the NLS1 selection. 

Recent optical spectroscopic surveys identified quasars with spectral shapes very similar to those of NLS1 in the distant universe. A quasar at $z = 6.62$ discovered from Pans-STARRS1 data \citep{tang2017,koptelova2017} exhibits high Ly$\alpha$ luminosity, narrow line width of $1300\pm 90 \mathrm{\ km\ s^{-1}}$ and fast UV flux variability on timescales of days, which are reminiscent of local NLS1s. Its Ly$\alpha$ luminosity accounts for 3\% of their total luminosity, which is a factor of two larger than those of other $z > 6.5$ quasars.

The Hyper Suprime-Cam (HSC) Subaru Strategic Program (SSP) discovered about 200 low-luminosity quasars at $z \gtrsim 6$ \citep[][and subsequent papers]{matsu2016}. Several objects in their sample have NLS1-like spectral properties such as a small broad Ly$\alpha$ line with weak continuum. \citet{onoue2019} reported near-infrared spectroscopic follow-up observations of six quasars from the HSC sample, including one object with an NLS1-like spectrum. The measured Mg\,\textsc{ii}-based BH mass of the NLS1 candidate is $\sim 10^{7.6} M_{\mathrm{\odot}}$, and it is accreting with an Eddington ratio ($\lambda_{\mathrm{Edd}}=L_{\mathrm{bol}}/L_{\mathrm{Edd}}$) of $1.1^{+0.5}_{-0.3}$. These findings further support the link to the NLS1 class. \citet{ayumi2024} measured the BH mass distribution of 131 HSC quasars at $z\sim 6$ utilizing the spectra of low-$z$ analogues selected from SDSS quasars. Their results show that quasars populating the low-mass end tend to show NLS1-like characteristics in their Ly$\alpha$ spectra. While it is not yet clear whether quasars with NLS1-like spectra (which we call NLQ1s in this study) represent the same population as classical NLS1s, NLQ1s may hold an important position for understanding the SMBH evolution, especially in the early growth phase of SMBH evolution. 

To advance the statistical studies of NLS1s that are missing in the $z>2$ regime, we search for distant NLQ1s in broad-line AGNs observed by the PFS-SSP survey, based on spectral information around Ly$\alpha$, and investigate their rest-UV properties. This paper describes the initial results of the NLQ1 search based on the PFS-SSP.

This paper is organized as follows. We describe the data and analysis in section 2. The main results are presented in section 3. The rest UV-properties of identified NLQ1s are discussed in section 4. We present a summary in section 5.
We adopt the cosmological parameters $H_0 = 70\ \mathrm{km\ s^{-1} Mpc^{-1}}, \Omega_{M}=0.3$, and $\Omega_\Lambda=0.7$. All magnitudes in the optical and near-infrared bands are presented in the AB system \citep{oke}.

\section{The PFS-SSP survey}\label{sec:2}


 \textquoteleft\={O}nohiʻula PFS is a massively multiplexed fiber spectrograph installed at the prime focus of the 8.2-meter Subaru telescope. This instrument covers optical and near-infrared wavelengths with 2386 science fibers distributed over the 1.25 deg field of view. The PFS enables the simultaneous observation of spectra with three arms of blue, red, and near-infrared. The spectral coverages and spectral resolutions of the blue, red, and near-infrared arms are 3800 -- 6500 \AA\, and $\sim 2.1$ \AA, 6300 -- 9700 \AA\, and $\sim 1.6$ -- $2.7$ \AA, 9400 -- 12600 \AA\, and $\sim 2.4$ \AA, respectively
\citep{tamura2016}. The PFS red arm has two resolution modes, and the data used in this study were acquired in the ``low-resolution'' mode with the resolving power of $\sim 3000$.

\textquoteleft\={O}nohiʻula PFS-SSP was launched to address key questions primarily in cosmology, galaxy and AGN evolution, and Galactic archaeology. The SSP survey is a large collaborative project with contributions from researchers in Japan, Princeton University, ASIAA in Taiwan, Caltech/JPL, Johns Hopkins University, Labora-
toire d’Astrophysique de Marseille (LAM), the Brazilian PFS
Participant Consortium, MPA/MPE in Germany, the Chinese
PFS Participant Consortium, and the North-East Participation
Group in the USA. The SSP observations are conducted across three main fields: (1) Cosmology, (2) Galactic archaeology, and (3) Galaxy Evolution (hereafter GE). \citet{takada2014} presents an overview of the framework. The PFS Science operations have been launched in March 2025 (S25A semester) and are scheduled for $\sim$360 nights over a 6-year period for the SSP runs. From March 2025 through July 2026, a total of 52.05 nights have been spent, of which observations were actually carried out on 38.25 nights.

\subsection{PFS-SSP Galaxy Evolution survey}
The GE survey is carried out to answer many important questions about (i) the timing, cause, and process of the intergalactic medium (IGM) reionization, (ii) the impact of the interplay between dark matter and baryons on galaxy evolution, and (iii) the evolutionary pathway of tight relations between fundamental galaxy properties. The GE survey will obtain $\sim 300$k spectra of galaxies at $z = 0.7$ -- $2.0$ to delineate their assembly history, obtain $\sim 40$k spectra of galaxies at $z = 2.1$ -- $3.5$ to reconstruct the large-scale HI gas distribution in the IGM, obtain $\sim22$k spectra of Lyman break galaxies at $z=3.5$ -- $7.0$ to studying early formation of galaxies and clustering strength, and detect $\sim 15$k Lyman-$\alpha$ emitters at $z = 2 .0$ -- $7.0$ to probe the state of the universe at the end of cosmic reionization. See \citet{jenny2022} for details about the science reasoning, and also see A. Goulding et al. (in preparation) for details about the survey design of the GE program.
 In the first PFS run semester, called S25A, the GE program observed targets in the ELAIS-N1 field, centered at RA $= 16^{\rm h}10^{\rm m}01^{\rm s}$, Dec $ = +54^{\circ}30'36''$ (J2000) \citep{oliver2000}, which served as the test field for the first year. The seeing conditions ranged from $\sim$ 0.4 to 1.2 arcsec, with a median value of 0.7 arcsec, and the range of total integration time is 2.0 -- 20 hours.

\subsection{Broad-line AGN targets}
The AGN targets were selected from data over a wide wavelength range from sub-millimeter/radio to X-ray. We expect that up to $\sim$30k AGN targets will be observed in the GE field, and up to $\sim 80$k AGN targets in the Cosmology field as ancillary targets. In this study, we use a sample of candidate broad-line (BL) AGNs observed in the GE field.
The BL-AGN candidates were selected from a photometric catalog referred to as the ``u2k catalog'' (version 2). This catalog is constructed by combining the HSC $g,i,r,z,y$-band photometry \citep{aihara2022}, $U$ ($u/u^*$)-band photometry provided by the Canada-France-Hawaii Telescope (CFHT) Large Area $U$-band Deep Survey (CLAUDS, \citealt{Sawicki2019}), $Y,J,H,K$-band photometry provided from UKIRT/WFCAM, and $Z,Y,J,H,K$-band photometry provided from VISTA/VIRCAM.
For the BL-AGN candidate selection, WFCAM was used for the $K$-band photometry. As the first criterion for target selection, we required the objects to be point sources with an $i$-band magnitude brighter than 24. The candidates were then selected using $u/u^*$-band and $K$-band excess at $z < 2.5 $, $u/u^*$-band dropout and $K$-band excess for $2.5 < z < 3.5$, the $g$-band dropout for $3.5 < z < 4.5$, the $r$-band dropout for $4.5 < z < 5.5$, and the $i$-band dropout for $5.5 < z < 6.5$, respectively. Further details of the target selection are described in Y. Toba et al. (in preparation).

\subsection{PFS spectra}
    The data reduction pipeline of the PFS is divided into the ``2D'' part and the ``1D'' part. In the 2D part, two-dimensional raw FITS data are received, and the reduction pipeline removes basic instrumental signatures from the raw data, applies wavelength calibration, and extracts one-dimensional spectra from each arm. After spectral extraction, sky-subtraction is performed.
    $\sim400$ fibers are allocated per pointing to sky, and 150 -- 200 fibers are allocated per pointing to flux standards, which accounts for $\sim25$\% of the fibers. The sky-background subtraction is performed by dividing the region on each camera into several blocks, computing the average flux from the sky fibers nearest to each science fiber, and subtracting this from the flux of the target fiber. Then, the pipeline merged three arm spectra into one spectral data, performed flux calibration, and finally combined spectra from multiple visits' spectra. The fully calibrated one-dimensional spectra, named ``pfsCoadd'', are delivered to users via Science Platform\footnote{https://hscpfs.mtk.nao.ac.jp/portal/\label{fn:SP}}, while users can also access the intermediate output files from the reduction and the results of quality assessment tests performed by the 2D pipeline. The 1D pipeline provides classifications of target types (i.e., galaxies, quasars, and stars), as well as redshift measurements for galaxies and quasars and radial velocities for stars. Additional key data products (e.g., FWHM and EW) will be provided in a forthcoming release of the 1D pipeline. \citet{tanaka2026} presents an overview of PFS science operation workflows. Since the reduction pipeline for the near-infrared arm (at $9400-12600$ \AA) is still under development, we restrict our analysis to the blue-arm and red-arm data in this study.

\section{Search for NLQ1s}

To identify NLQ1s, we first cross-matched the target list of the BL-AGN candidates with observed objects in the GE field within 1.0 arcsec. 105 targets were observed in the S25A GE runs. 
We downloaded their pfsCoadd spectra (pipeline version: \texttt{w2026.06}, release name: \texttt{S25A\_April2026}) from the Science Platform\footnotemark[\ref{fn:SP}]. We used systematic redshifts measured by the PFS 1D pipeline for our analysis. For spectra whose pipeline redshifts were not correct based on visual inspection, we re-measured the redshift using the central wavelength of the detected Ly$\alpha$ peak. The Ly$\alpha$ identification was verified using different criteria depending on redshift; for low-$z$ objects, by the presence of other emission lines; for high-$z$ objects, by the flux decrement blueward of the line and its asymmetric profile. The objects for which the pipeline redshifts were wrong span $z=2.5$ -- $3.6$, at which the severe absorption of the blue half of the Ly$\alpha$ profile, as typically seen in higher-$z$ quasar spectra, is not expected to be significant.

In order to find NLQ1s at high redshift, we used Ly$\alpha$ emission lines at $z=2.2$ -- $6.9$, which fall in to the PFS spectral coverage of 4000 - 9600 \AA. Within this redshift range, 57 BL-AGN targets were found.

\subsection{Spectral fitting}
We search for NLQ1s among these 57 BL-AGN targets, which are hereafter called the PFS quasar sample.
We obtained best-fit continuum models and Ly$\alpha$ emission line widths using the publicly available \texttt{PyQSOFit} code developed by \citet{guo2018}.

All spectra were corrected for Galactic extinction using the dust map of \citet{schlegel1998}, and the extinction law of \citet{fitzpatrick1999} with $R_V=3.1$.

By comparing with the actual scatter observed in the spectra, a fraction of the PFS quasars have underestimated errors in the pfsCoadd spectra. Therefore, we corrected the error spectra provided by the pipeline in the following manner. We assume that (1) the noise spectral shapes are correct and only their global amplitudes are incorrect, (2) the error distributions are described by a Gaussian, (3) the continuum windows used below have no signals of emission line, absorption line, or IGM. Since our main interest is in the Ly$\alpha$ and C\textsc{iv} lines, we corrected the noise amplitudes using continuum regions around these emission lines  (1280 -- 1290, 1315 -- 1360, 1425 -- 1470, 1680 -- 1710, and 1975 -- 2050\,\AA). The corrected noise ($\sigma_i^{\mathrm{true}}$) is defined as follows.
\begin{equation}
    \sigma^{\mathrm{true}}_i = \alpha \cdot	\sigma_i^{\mathrm{pipe}}
\end{equation}

\noindent where $\sigma_i^{\mathrm{pipe}}$ is the noise estimated in the pipeline, $i$ denotes the pixel index, and $\alpha$ is the scaling factor. In the continuum region, where the signal varies smoothly, flux fluctuations can be attributed to noise. We compute the ratio of the standard deviation of the residual spectrum -- obtained by subtracting the mean flux from the observed flux in the continuum region -- to the error spectrum, which gives an estimate of $\alpha$. We applied this correction to objects for which the pipeline noise differs from the noise directly determined from the observed spectra by more than $\pm 20\%$.

Since the pipeline background subtraction is still being refined, we applied an additional correction to the flux level using HSC single-band photometry covering rest-frame 1350 \AA\, at which BH masses are estimated later. We convolved the PFS spectra with the HSC filter transmission function to compute representative PFS flux within the band. We then calculated the flux ratio between the PFS measurement and HSC photometric measurement, and corrected the flux deviation accordingly. This correction is not applied to objects with negative median representative flux.

To mitigate the effects of the Ly$\alpha$ forest and damped Lyman-alpha system (DLA) absorption, we masked those pixels blueward of Ly$\alpha$ and with fluxes less than 10\% of the median continuum flux at $\lambda =1275$ -- $1290$\ \AA\ for $z \le 4$ targets. We excluded the data blueward of the Ly$\alpha$ peak from the fitting for targets at $z>4$.

We then performed the spectral fitting analysis on the processed observed spectra. The continuum was fit within the line-less windows  (table \ref{tab:continuum2}) across the entire wavelength region ($3800$ -- $9700$ \AA\, in the observed frame).

\begin{table}[t]
\centering
\caption{Rest-frame continuum windows for spectral fitting}
\label{tab:continuum2}
\begin{tabular}{c}

\hline\hline
Range [\AA] \\
\hline
1 : 1275--1290\\
2: 1350--1360\\
3: 1445--1465 \\
4: 1690--1705 \\
5: 1770--1810 \\
6: 1970--2400 \\
7: 2480--2675 \\
8: 2925--3400 \\

\hline
\end{tabular}
\end{table}

The continuum model is described as:

\begin{equation}
    f_{\rm cont}(\lambda) = f_{\rm power-law}(\lambda) + f_{\rm poly}(\lambda) + f_{\rm{Fe}\,\textsc{ii}}(\lambda)
\end{equation}

\noindent where the power-law continuum $f_\mathrm{power-law}$ represents the accretion disk emission, which is expressed as

\begin{equation}
f_{\mathrm{power-law}}(\lambda) = a_0(\lambda/\lambda_0)^{a_1},
\end{equation}

\noindent with a reference wavelength $\lambda_0=3000$\,\AA. $\alpha _0$, $\alpha_1$ are free parameters. 

\begin{figure*}[!t]
    \centering
    \includegraphics[width=18cm]{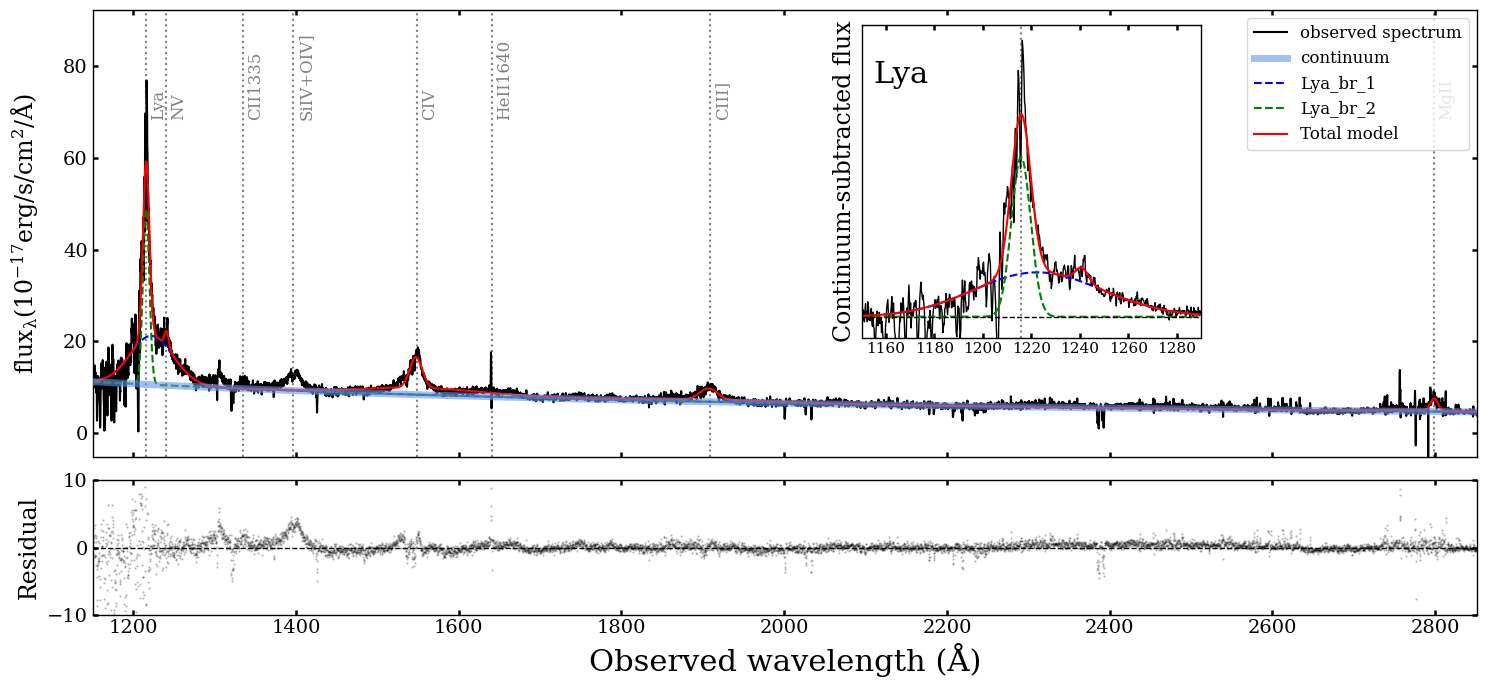}
    \caption{An example of spectral fitting for a z = 2.4 PFS quasar spectrum. We show the observed data in the rest-frame with a black solid line. The semi-transparent blue line shows the continuum model. 
    The blue and green dashed lines show the best-fit broad components in Ly$\alpha$. The red line shows the best-fit model of the combined continuum and emission line components. The inset displays the Ly$\alpha$ region after continuum subtraction. The vertical dotted lines indicate the expected central wavelengths of emission lines, including Ly$\alpha$, N\,\textsc{v}, C\,\textsc{ii}, Si\,\textsc{iv}+O\,\textsc{iv}], C\,\textsc{iv}, He\,\textsc{ii}, C\,\textsc{iii}], and Mg\,\textsc{ii}. The black dashed lines show the zero flux level. Residuals are also shown at the bottom.
    {Alt text: An example plot of spectral fitting on top. The vertical axis shows flux density in units of 10 to the power of negative 17 erg per second per square centimeter per angstrom. The horizontal axis shows observed wavelengths in units of angstroms. Residual plots are also shown on the bottom.}}
    \label{fig:qsofit1}
\end{figure*}

\noindent $f_\mathrm{Fe II}$ represents multiple Fe\,\textsc{ii} emission-line blends, which produce a complex pseudo-continuum in the optical–UV quasar spectra.

\begin{equation}
    f_{\mathrm{Fe\,II}}(\lambda) = c_0 F_{\mathrm{Fe\,II}}(\lambda,\, c_1,\, c_2)
\end{equation}
\noindent 
The Fe\,\textsc{ii} template is scaled and shaped by three parameters ($c_i$ ) that are fit to each spectrum. $c_0$, $c_1$, and $c_2$ are the normalization factor, the FWHM of the Gaussian convolution kernel, and the wavelength shift, respectively.
At 1000 -- 2200\,\AA, we used the modified UV Fe\,\textsc{ii} template \citep{shen2019} that is originally constructed by \citet{vestergaard2001}. The templates from \citet{salviander2007}, and \citet{tsuzuki2006} were used for 2200 -- 3090\,\AA\, and for 3090 -- 3500\,\AA, respectively. We used the optical Fe\,\textsc{ii} template based on T. Boroson et al. (1992). All the templates are built into \texttt{PyQSOFit}.

\noindent Some of the quasar spectra are affected by dust reddening, which causes a deviation from a simple power-law continuum shape. We account for this effect by incorporating a polynomial function into the continuum model, denoted $f_{\mathrm{poly}}$, with the free parameters $b_i$.


\begin{equation}
f_{\mathrm{poly}}(\lambda) = \sum_{i=1}^{n} b_i(\lambda - \lambda_0)^i,
\end{equation}
\noindent where $\lambda_0$ is the reference wavelength and $n$ is the polynomial order. Here, we adopt a third-order polynomial, i.e., $n=3$.

We do not include the host-quasar templates since we are targeting quasars beyond $z=2.2$ and the observational sensitivity is insufficient to detect the host galaxy. 
The emission line profiles were modeled using up to three Gaussian components for Ly$\alpha$, one for N\,\textsc{v}, two broad-only components for C\,\textsc{iv}, up to two broad plus one narrow components for C\,\textsc{iii}] and for Mg\,\textsc{ii}. At $z \gtrsim 6$, for objects where the signal-to-noise ratio is insufficient to fully separate Ly$\alpha$ and N\,\textsc{v}, we do not perform a fit to N\,\textsc{v}. In addition, Fe\,\textsc{ii} emission components and the polynomial component were excluded from the continuum fitting. We set the threshold for the broad component at a minimum FWHM of $\mathrm{1000\, \ km\,s^{-1}}$. 
All uncertainties are estimated by the Monte Carlo approach built into \texttt{PyQSOFit}. 
Figure \ref{fig:qsofit1} shows an example of spectral fitting for one PFS quasar spectrum.

\subsection{Identification of NLQ1s}

\begin{figure*}[!h]
 \begin{center}
    \includegraphics[height=20cm, width=18cm]{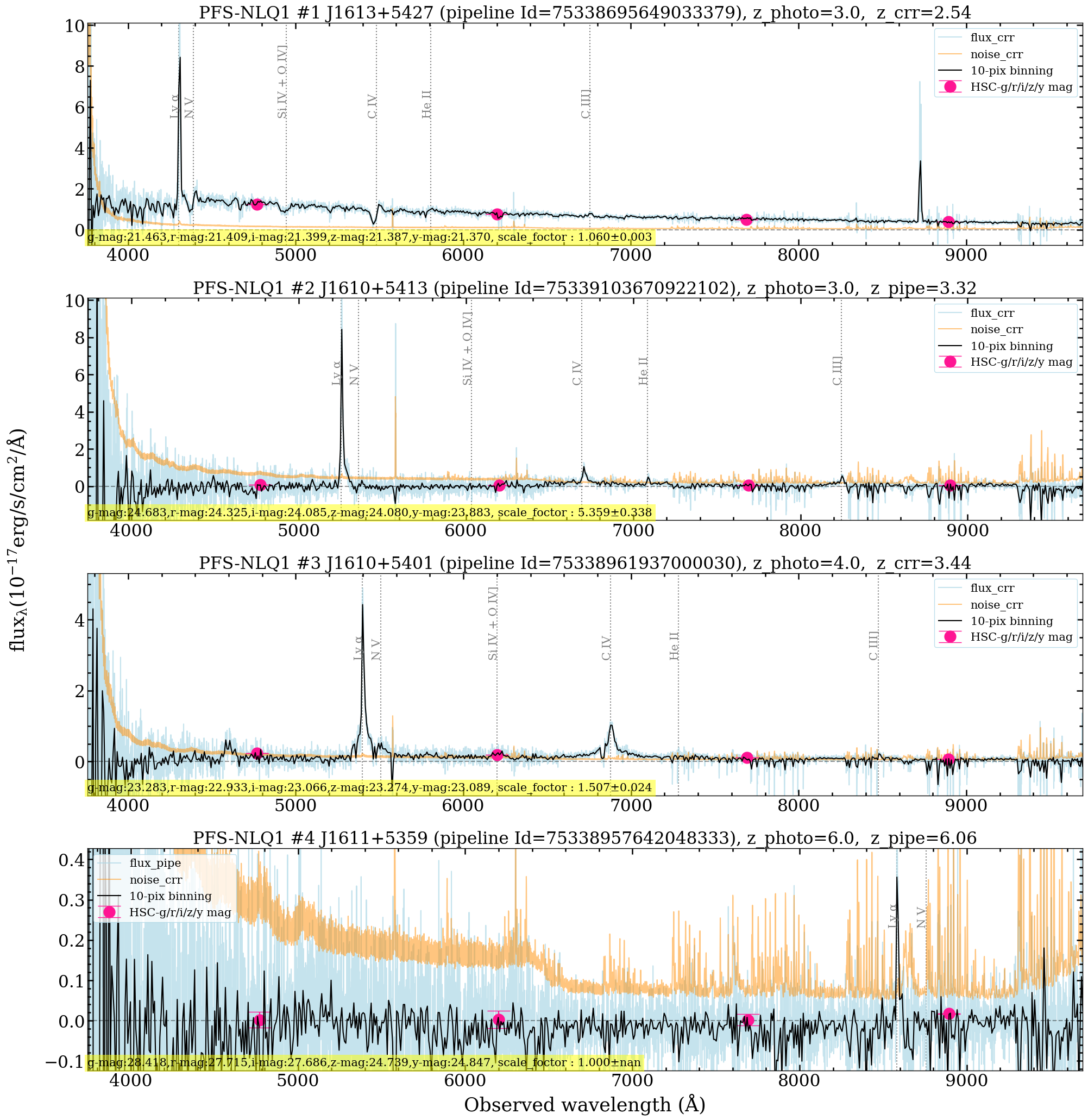} 
    \caption{ PFS spectra of the four NLQ1s. Blue lines show unbinned spectra, and black lines show 10-pixel binning spectra. The spectrum of J1611+5359 is not the coadded spectrum obtained directly from the pipeline, but a custom coadded spectrum constructed using two prestacked spectra. Four HSC band magnitudes are shown with pink dots. Except for J1611+5359, the flux of each object has been corrected with a single HSC band magnitude. The values of scale factors for flux correction and of HSC magnitudes are shown in the yellow box at the lower left. Orange lines represent corrected flux errors. The vertical dotted lines indicate the expected central wavelengths of emission lines, including Ly$\alpha$, N\,\textsc{v}, Si\,\textsc{iv}+O\,\textsc{iv}], C\,\textsc{iv}, He\,\textsc{ii}, C\,\textsc{iii}]. Target name, pipeline ID, and redshift are shown at the top of each panel. Half of the objects use spectroscopic pipeline redshift ($z$\_$\rm{pipe}$), and the other half use the remeasured redshift ($z$\_$\rm{crr}$) estimated from the Ly$\alpha$ emission line peak. The red arm covers up to 9700\,\AA. The apparent line at $\sim 8750$\,\AA\ in the spectrum of J1613+5427 is due to a cosmic ray.
    {Alt text: A plot with four panels, each showing spectral flux in 10 to the power of negative 17 erg per second per square centimeter per angstrom on the vertical axis and wavelengths in angstroms on the horizontal axis.}  }
    \label{fig:spec_nls1}
\end{center}
\end{figure*}

We searched for $z > 2$ NLQ1s in the observed PFS quasars with the requirement of Ly$\alpha$ FWHM $< 2000\rm{\ km\ s^{-1}}$. The Ly$\alpha$ FWHM is derived numerically from the composite profile of all Gaussian components used in the fit, whose sigma exceeds 0.0017 in log-wavelength space in units of \AA. We corrected the line widths for instrumental broadening. Objects with no detected broad lines in any of the emission lines are classified as Type-2 quasars (narrow-line AGNs) \citep[e.g.,][]{alex2013}. 
From the fitting for the 57 BL-AGN targets, seven targets meet the above FWHM requirements. Three of these objects either showed no detectable broad line, or a poor fit. Excluding these three objects, we identified three NLQ1s at $z = 2.54$ -- $3.44$ and one promising candidate at $z = 6.06$ (figure \ref{fig:spec_nls1}). As the pfsCoadd of the $z=6.06$ candidate exhibited an artificial flux jump between the blue and red arms, we removed 4 of the 6 prestacked spectra that show the flux jump and created a custom coadded spectrum with an inverse-variance weighted mean method, which we use for the following analysis. Their Ly$\alpha$ properties with some key information for each object are listed in table~\ref{tab:linepro}. J1611+5359 is one of the most distant NLQ1 candidates ever known. We checked the prestacked spectra of this target because only one emission line at $\sim 8600 $ \AA\, is detected. 
We confirmed that the asymmetric emission line is present in all prestacked spectra (figure \ref{fig:prestack} in Appendix 1), and its HSC colors are also in good agreement with those of $z \sim 6$ quasars.

The measured Ly$\alpha$ FWHM of the four NLQ1s range from $1410\mathrm{\ km\ s^{-1}}$ to $1510 \mathrm{\ km\ s^{-1}}$. They have significantly broad components in Ly$\alpha$, and are clearly distinguished from Type-2 quasars. J1613+5427 and J1610+5401 show blue continua, which also supports that these are definitely Type-1 quasars. The NLQ1s show a prominent Ly$\alpha$ line on extremely low continuum, which is quite different from the spectral shapes of typical quasars. The EW (Ly$\alpha$) measurements range from 18 to 313 \,\AA, with an average of  $\sim 190$ \,\AA, which are significantly larger than the typical value of 93 \,\AA\, in quasars, as reported in \citet{vandenberk2001}. Although the selection of NLQ1s in this study was based only on the line widths, our search revealed that these objects also exhibit remarkably large EW values.

We also calculated the Ly$\alpha$ luminosities ($L_{\mathrm{Ly}\alpha}$) of the NLQ1s. The NLQ1s have a median luminosity of $\log (L_{\mathrm{Ly}\alpha}/[\mathrm{erg\,s^{-1}}]) \sim 43.89$. For the three objects other than the $z=6$ source, rest-frame 1350\, \AA\ falls within the wavelength coverage of the red arm, allowing us to derive the bolometric luminosity using a bolometric correction at 1350\, \AA\ (see section 4.1\label{subsec:BHmass} for details). The ratio of the Ly$\alpha$ luminosity to the bolometric luminosity of the NLQ1s ranges from $\sim 0.3$ to $4.0$\%, with a median of $\sim 3.1$\%.

 For J1611+5359 at $z=6$, the red side of Ly$\alpha$ is absorbed by the IGM damping wing, making the broad component unconstrained. We therefore expect that this target has relatively large errors for these measurements, while we do not use the blueward of Ly$\alpha$ in the spectral fitting.

 We show the results of the spectral fitting around Ly$\alpha$ for the four NLQ1s in figure \ref{fig:qsofit2} (appendix 2).

\clearpage

\clearpage

\clearpage

\clearpage


\clearpage
\pdfpageattr{/Rotate 90}
\thispagestyle{empty}

\noindent\makebox[\textwidth][c]{%
\rotatebox{90}{%
  \begin{minipage}{0.95\textheight}
  \centering
    \captionof{table}{
        Ly$\alpha$ and other basic properties of the NLQ1s.
    }
    \label{tab:linepro}

    \renewcommand{\arraystretch}{2.}
    \setlength{\tabcolsep}{4pt}
    \resizebox{\linewidth}{!}{%
    
    \begin{tabular}{lcccccccccccccc}
    \hline
    (1) & (2)& (3) & (4) & (5) & (6)& (7) & (8) & (9) & (10) & (11)& (12) & (13) & (14) & (15) \\
    Name & pipeline ID & R.A. & Decl.& $g_{\mathrm{AB}}$ & $r_{\mathrm{AB}}$ & $i_{\mathrm{AB}}$ & $z_{\mathrm{AB}}$ & $y_{\mathrm{AB}}$ & redshift & Exp. Time & seeing &$\mathrm{FWHM_{Ly\alpha}}$ & $\mathrm{EW_{Ly\alpha}}$ & $\log L_\mathrm{{Ly\alpha}}$ \\
     & & & & (mag) & (mag) & (mag) & (mag) & (mag) &  & (hr)& (arcsec) & ($\mathrm{km\,s^{-1}}$) & (\AA) & ($\mathrm{erg\,s^{-1}}$) \\
    \hline
    J1613+5427 & 75338695649033379 & 243.472356	&54.462906 &  21.46  & 21.41 & 21.40  & 21.39  & 21.37  & 2.54 & 2.6 & 0.81& 1410 $\pm$ 1 & 18 $\pm$ 1 & 43.668 $\pm$ 0.010 \\

    J1610+5413 & 75339103670922102 & 242.630938 & 54.222391 &  24.68  & 24.33  & 24.09  & 24.08  & 23.88 & 3.32 & 16.5 & 0.72 & 1409 $\pm$ 5 & 256 $\pm$ 5 & 44.193 $\pm$ 0.011 \\
    
    J1610+5401 & 75338961937000030 & 242.746705 & 54.018550 &  23.28  & 22.93  & 23.07  & 23.27  & 23.09  & 3.44 & 4.0 & 0.62 & 1510 $\pm$ 7  & 170 $\pm$ 4 & 44.041 $\pm$ 0.005 \\
    
    J1611+5359 & 75338957642048333 & 242.819272 & 53.986159 &  28.42  & 27.72  & 27.69  & 24.74  & 24.85  & 6.06 & 1.5 & 0.59 & 1410 $\pm$ 61 & 313 $\pm$ 31  & 42.958 $\pm 0.260$  \\
    
    \hline

    \end{tabular}
    }
    \vspace{2mm}
    \begin{minipage}{\linewidth}
    {\footnotesize
    \textbf{Note.} Columns (1)-(4) give the object names, pipeline IDs, and coordinates. The pipeline ID is referred to as \texttt{objectId} in the PFS pipeline.
    Columns (5)--(9) show the HSC magnitudes.
    Columns (10), (11), and (12) list the spectroscopic redshifts, integrated time, and median seeing, respectively. Columns (13)-(15) show the measurements of Ly$\alpha$ lines.
    The line widths are corrected for instrumental broadening.}
    \end{minipage}
    
\end{minipage}%
}%
}
\clearpage
\pdfpageattr{}

\clearpage

\section{Rest-UV properties of the NLQ1s}

\subsection{Black hole mass measurements}
\label{subsec:BHmass}
In the following, we discuss the BH masses of the NLQ1s. For comparison, we also measure the BH masses of the other PFS quasars.

23 objects among the PFS quasars at $2 < z < 6$ show detection of C\,\textsc{iv} emission line, which we use to measure their BH masses. We were able to measure the masses for only three of the PFS NLQ1s.

The BH masses are estimated from the single epoch method with the following empirical relation:

\begin{equation}
 \mathrm{log}\frac{M_{\mathrm{BH}}}{M_{\odot}} = 0.66 + 0.53\ 
 \mathrm{log}\frac{\lambda{L_{1350}}}{10^{44}\mathrm{erg^{-1}}}+2\ 
 \mathrm{log}\frac{\mathrm{FWHM}}{\mathrm{km\ s}}
\end{equation}
\\

\noindent where FWHM is the line width of C\,\textsc{iv} emission line. $\lambda L_{1350}$ is the monochromatic luminosity at 1350\,\AA, radiated from the accretion disk. This equation has been built on the relation between the radius of the line-emitting region and the continuum luminosity in the local universe \citep[e.g.,][]{kaspi2005}. The scaling factors (0.66, 0.53) are employed in the measurement of BH mass via C\,\textsc{iv} \citep{vestergaard2006}. The continuum luminosity values obtained by the \texttt{PyQSOFit} were used for most of the targets. The $L_{\rm{1350}}$ of  J1610+5413 (NLQ1) was found to be negative. This is likely attributable to systematics in the pipeline reduction. We therefore replaced the flux at rest-frame 1350 \AA\  with the photometric measurement of the HSC $r$-band for this exceptional case. Since the flux uncertainties of the observed spectra currently have large uncertainties, and it is difficult to rigorously predict their magnitude, we conservatively adopted a uniform 30\% uncertainty of $L_{1350}$ whenever the fitting-based measurement uncertainty was smaller than 30\% of $L_{1350}$. For all objects, we measured the C\,\textsc{iv} line widths using
our own code applied to the spectra (i.e., not from the PFS 1D pipeline or \texttt{PyQSOFit}), but the continuum flux was subtracted using the best-fit model estimated by \texttt{PyQSOFit}. In the line-fitting, we visually inspected and masked the absorbed regions where present, and adjusted the number of Gaussian components. The C\,\textsc{iv} FWHMs were measured from the total line profile of the one- or two-Gaussian fit. We corrected the line widths for
instrumental broadening. The quality assessment plots of the line fitting for the NLQ1s are shown in Figure \ref{fig:civ-fit1} (Appendix 3). The associated uncertainty was derived using the MCMC method. The BH mass uncertainties were derived by propagating the measurement uncertainties of the luminosities and line widths.

With the BH mass ($M_{\mathrm{BH}}$) determined, we can also calculate the Eddington luminosity using the following formula:

\begin{equation}
    L_{\mathrm{Edd}} = 1.26\times10^{38}(\frac{M_{\mathrm{BH}}}{M_{\odot}})\ \mathrm{erg\ s^{-1}}
\end{equation}

We also estimated the bolometric luminosity ($L_{\mathrm{bol}}$) by applying a bolometric correction to the monochromatic luminosity with $L_{\rm bol} =3.81 \times \lambda L_{1350}$ \citep{richard2006}, where $L_{1350}$ is a monochromatic luminosity at 1350\,\AA. The measured values used for the BH mass estimation, bolometric luminosities, and Eddington ratios ($\lambda_{\rm{Edd}}=L_{\rm bol}/L_{\rm Edd}$) are listed in table \ref{tab:properties_civ}. 


\begin{table*}[t]
\centering

\caption{Around C\,\textsc{iv} properties of available PFS-NLQ1s.}
\label{tab:properties_civ}

\newlength{\civtablewidth}
\renewcommand{\arraystretch}{1.1}
\setlength{\civtablewidth}{\textwidth}
{\centering
\resizebox{\civtablewidth}{!}{%

\begin{tabular}{lcccccc}
\hline\hline
Name &
log $\lambda L_{1350}$ &
log $L_{\rm bol}$ & FWHM$_{\rm C\,\textsc{iv}}$ &
log $M_{\rm BH,\,C\,\textsc{iv}}$  & $\rm \lambda_{Edd}$ & Flag\_BAL\\
 &
($\rm erg~s^{-1}$) &
($\rm erg~s^{-1}$) & 
($\rm km~s^{-1}$) &
($M_{\odot}$) &
\\
\hline
J1613+5427 & 45.54 $\pm$ 0.13 & 46.12 $\pm$ 0.71 &  774 $^{+302}_{-267}$ & 7.26$_{-0.31}^{+0.35}$   & 5.88 $^{+10.48}_{-10.70}$& Yes\\
J1610+5413 & 44.09 $\pm$ 0.13 & 44.67 $\pm$ 0.71 & 1308 $^{+106}_{-107}$ & 6.94 $_{-0.10}^{+0.10}$  & 0.64 $^{+1.05}_{-1.54}$ & No \\
J1610+5401 & 44.98 $\pm$ 0.13 & 45.56 $\pm$ 0.71 & 1723 $^{+83}_{-73}$ & 7.65 $_{-0.08}^{+0.08}$  & 0.42 $^{+0.70}_{-0.70}$ & No\\

\hline
\end{tabular}%

}
\par}

\begin{minipage}{\civtablewidth}
{\footnotesize
\textbf{Note.} The line widths are corrected for instrumental broadening.}
\end{minipage}

\end{table*}

\begin{figure}[!t]
\includegraphics[height=9cm, width=8.5cm]{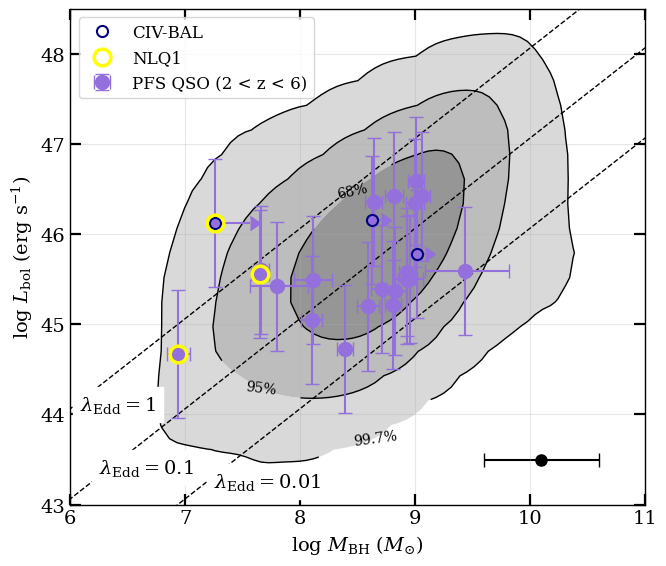} 
\caption{Distribution of quasars in the BH mass and bolometric luminosity space. The PFS sample is shown with purple dots, and the NLQ1s are shown with yellow open circles. The PFS quasars with C\,\textsc{iv} broad absorption lines are also shown with dark blue circles. The black contours correspond to the SDSS-DR16 quasars compiled by \citet{wu2022}. The diagonal lines show the constant Eddington ratios of $L_{\mathrm{bol}}/L_{\mathrm{Edd}}$=1, 0.1, 0.01 from top left to bottom right. The typical systematic uncertainty of the $M_{\mathrm{BH}}$ measurements (0.5 dex) is shown with the error bar at the lower right.
{Alt text: Black hole masses of our samples plotted against the bolometric luminosities. }}
\label{fig:dist_MBH}
\end{figure}

Figure \ref{fig:dist_MBH} shows the BH mass distribution, including other PFS quasars. J1613+5427 has log$(M_{\mathrm{BH}}/M_{\odot})\sim 7.26$ with $\lambda_{\mathrm{Edd}}\sim 5.87$. We should note that the C\,\textsc{iv} profile of J1613+5427 is strongly affected by a BAL trough; although we performed the fitting carefully by masking the absorbed region, there is a possibility that the C\,\textsc{iv} profile is not accurately reproduced. The BAL trough potentially leads to an underestimate of the BH mass and an overestimate of the Eddington ratio. J1610+5401 has log$(M_{\mathrm{BH}}/M_{\odot})\sim 6.94$ with $\lambda_{\mathrm{Edd}}\sim 0.42$. J1613+5427 has slightly larger BH mass of log$(M_{\mathrm{BH}}/M_{\odot})\sim 7.65$ compared to other NLQ1, and accreting with $\lambda_{\mathrm{Edd}}\sim 0.64$. The NLQ1s have a median logarithmic BH mass of $7.25^{+0.37}_{-0.29}$, which is lower by $\sim 1.5$ dex than the value of $8.79^{+0.09}_{-0.11}$ for the other PFS quasars. The median Eddington ratio of the NLQ1s is $\lambda_{\mathrm{Edd}}=0.72^{+1.26}_{-0.81}$, while that of other PFS quasars is $\lambda_{\mathrm{Edd}}=0.05^{+0.03}_{-0.02}$. The median bolometric luminosity of the three NLQ1s is log($L_\mathrm{bol}/[\mathrm{erg\, s^{-1}}]$) $= 45.48^{+0.62}_{-0.65}$, which is consistent with the other PFS quasars within the uncertainties (log($L_\mathrm{bol}/[\mathrm{erg\, s^{-1}}]$) $= 45.67^{+0.25}_{-0.24}$). 
The above uncertainties were estimated using the bootstrap method. These results indicate that, despite there being no significant difference in the bolometric luminosities between NLQ1 and the other PFS quasars, the BH masses and Eddington ratios of the NLQ1 differ significantly from those of the normal quasars. Similarly to NLS1s, NLQ1s may be characterized more by their BH masses and accretion rates than by their luminosities.


\citet{Rakshit2021} (R21) measured the BH masses and Eddington ratios of 2684 NLS1s at $z = 0.8-2.5$. Many of their BH masses are obtained by C\,\textsc{iv} lines. The R21 NLS1 sample have median log($M_{\mathrm{BH}}/M_{\odot}$) of $8.01 \pm 0.35$ with $\mathrm{\lambda_{Edd}\sim 1.0}$ and the median log($L_{\mathrm{bol}}/[\mathrm{erg\, s^{-1}}]$) is $46.00 \pm 0.42$. We find that the median BH masses of the R21 sample and our NLQ1 are broadly consistent with one another within the measurement uncertainties ($\sim 1.6\,\sigma$), indicating no statistically significant difference between the two samples. Our bolometric luminosities are also comparable to those of R21. The two samples occupy a similar region in the $M_{\mathrm{BH}}$-luminosity plane.

\begin{figure}[!t]
\includegraphics[height=8cm, width=8cm]{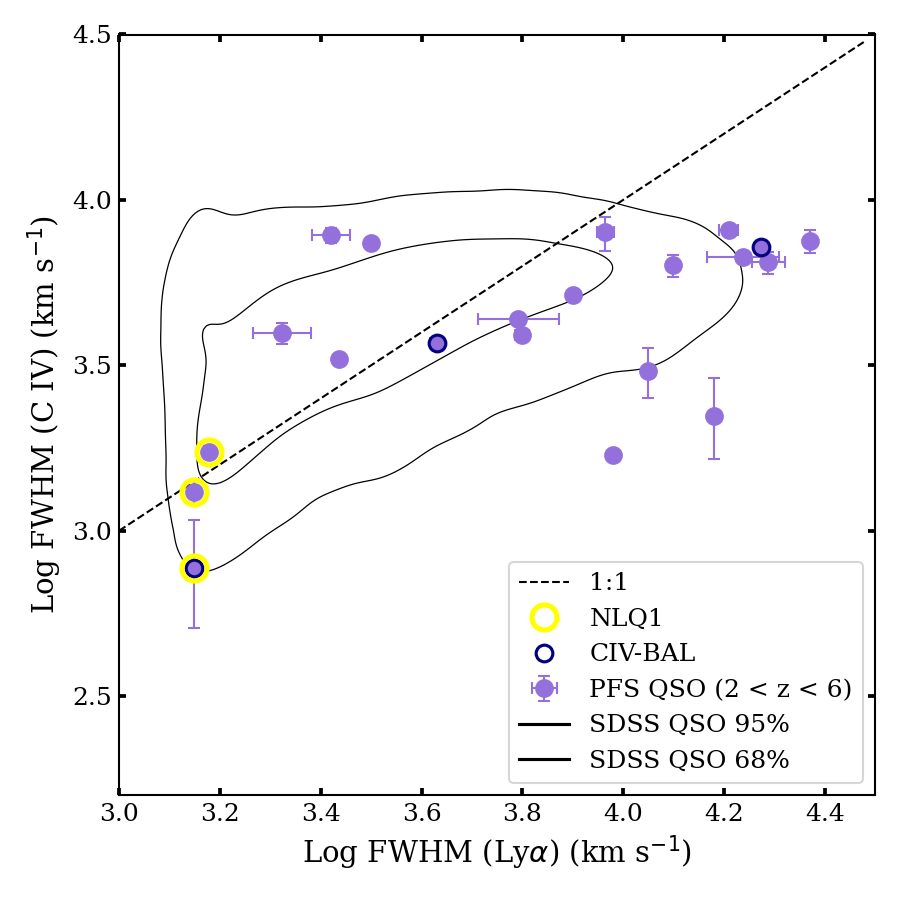} 
\caption{Comparison of the FWHM of Ly$\alpha$ and C\,\textsc{iv}, measured from the total line profile of the multi-Gaussian fit. Purple points show the PFS sample. The PFS NLQ1s are highlighted as yellow circles. Dark blue circles represent the BAL quasars. Gray contours show the SDSS sample \citep{wu2022}.
{Alt text: Comparison plot of Carbon IV line widths versus Lyman-alpha line widths.}}
\label{fig:fwhm_lya_vs_civ}
\end{figure}

Ly$\alpha$ has complex radiative transfer processes and is therefore difficult to interpret. The spatial relation between the emission regions of Ly$\alpha$ and C\,\textsc{iv} --- the latter of which is thought to arise from gas orbiting in the immediate vicinity of the BH --- is not straightforward. Nevertheless, when we compare the FWHMs of Ly$\alpha$ and C\,\textsc{iv}, the two non-BAL PFS NLQ1s lie close to the one-to-one line (figure \ref{fig:fwhm_lya_vs_civ}). 
The correlation between the Ly$\alpha$ and C\,\textsc{iv} line widths may suggest that, because the
overall extent of the BLR is compact while individual gas clouds are optically thick, the regions where C\, \textsc{iv} and Ly$\alpha$ are efficiently emitted overlap substantially in space. In addition, combined with the high EW results, this suggests that the line-emitting gas has a relatively large column density and efficiently converts the ionizing energy output from the accretion disk into line photons. This interpretation is consistent with the non-negligible outflow fraction among our NLQ1s discussed in the next subsection.

We note that the bolometric correction is based on the average AGN SED, and \citet{richard2006} indicates that a 50\% uncertainty is associated with each luminosity measurement. Moreover, for rare objects like those studied here, which show exceptionally weak continuum emission and strong emission lines, the actual uncertainty is likely considerably larger.

\subsection{Absorption features}

A fraction of quasars exhibit broad absorption line (BAL) trough extending over more than 2000 $\mathrm{km\ s^{-1}}$ \citep{weymann1991}. This can be related to one of the quasar activities, i.e., outflowing gas, which has the potential to drive the AGN feedback. The BAL fractions are $\sim 10-50\%$ in UV and optical survey data \citep[e.g.,][]{Foltz1990, trump2006,Paris2018,bischetti2023} depending on redshift, while some studies imply that the BAL fraction is strongly affected by the sample selection method \citep[e.g.,][]{Dai2008}. 
The similarity of optical properties between NLS1 and BAL quasars has been proposed by several studies \citep[e.g.,][and references therein]{brandt2000,Boroson2002,grupe2015}. In addition, \citet{Lawrence1998} and \citet{xu2007} indicate that NLS1s have low-density narrow line regions, and the gas has a tendency to be easily blown away by the outflowing wind. A local NLS1 galaxy, WPVS007, which shows a rapid decline in X-ray luminosity, has mini-BAL and BAL features in its UV spectrum \citep{leighly2009}. 

One PFS NLQ1, J1613+5427, likely has broad absorption features at the wavelengths blueward of its C\, \textsc{iv} and N\,\textsc{v}. 
Here, we measure the strength of the absorption lines in C\,\textsc{iv} in order to provide a more quantitative assessment.

\begin{figure}[!t]
 \begin{center}
\includegraphics[height=4cm, width=5cm]{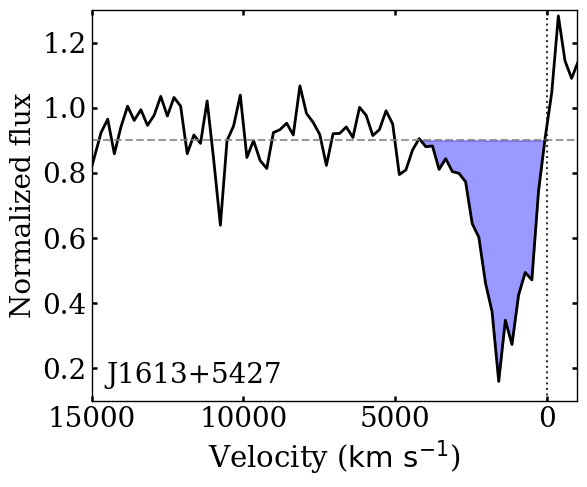} 
 \end{center}
\caption{The continuum-normalized spectra around C\,\textsc{iv} of J1613+5427, which is smoothed with 5 pixels. Blue highlight shows the absorption line region. Horizontal dashed line represents the normalized flux of 0.9, which is the threshold for AI, and vertical dotted line represents the wavelength of C\,\textsc{iv} ($\lambda1549$). {Alt text: Spectral plot showing the Carbon IV absorption line in J1613plus5427, with the absorption region highlighted and a threshold line at normalized flux 0.9.}}
\label{fig:spec_bal}
\end{figure}

 BAL quasars have been identified via the ``balnicity index '' (BI) defined in \citet{weymann1991}; it calculates blueshift velocities in the range of 3000 to 25000  $\mathrm{km\ s^{-1}}$ relative to the emission line center. BI considers not only the width but also the depth of absorption. We calculated the BI for J1613+5427 and found that its absorption does not meet the BI criterion. Another indicator ``absorption index'' (AI) proposed in \citet{Hall2002} uses a more relaxed requirement for detecting absorption troughs as follows.

\begin{equation}
    \mathrm{AI}
    = - \int_{25000}^{0}
    \left[ 1 - \frac{f(v)}{0.9} \right] C(v)\, \mathrm{d}v
    \label{eq:AI}
\end{equation}

\noindent where $f(v)$ is the normalized flux density as a function of velocity displacement from the center of C\, \textsc{iv} emission line, and $C(v)=1$ is defined as follows: $C(v) = 1$ if the absorption trough continuously extends for more than $450 \mathrm{km\ s^{-1}}$, and $C(v) = 0$ otherwise. The integration is performed over the velocity range from 0 to 25000 $\mathrm{km\ s^{-1}}$ blueward of the emission line center. Figure \ref{fig:spec_bal} presents the region of the absorption trough. This NLQ1 has AI $=1219 \ \mathrm{km\ s^{-1}}$ with the simple velocity width of $\mathrm{3495\ km\ s^{-1}}$ corresponding to a contiguous region where the flux is depressed by more than 10\% relative to the continuum, which is classified as a mini-BAL quasar.
The C\,\textsc{iv} BAL fraction among the PFS NLQ1s was found to be $\sim 33$\%, derived from three objects excluding the $z = 6$ NLQ1, whose C\,\textsc{iv} line falls in the near-infrared arm. This BAL fraction is close to the AI-based BAL fraction ( $\sim 27$\%) found for the SDSS DR14 quasars \citep{Paris2018}, despite the UV absolute magnitude of the NLQ1s being about 3.7 mag fainter than that of SDSS DR14 quasars at $z = 2.5$ -- $3.4$.
The properties of J1613+5427 are not limited to the mini-BAL features: it also exhibits a notable blue continuum compared to other identified NLQ1s. A relatively developed accretion disk may be blowing away the surrounding gas, giving rise to a mini-BAL wind, which may partially obscure the central source. On the other hand, it is known that BAL quasars tend to have a redder continuum than non-BAL quasars \citep[e.g.,][]{trump2006, gibson2008}. Reddening is quantified as the ratio between the continuum flux densities, defined as 

\begin{equation}
R_{\rm{red}}\equiv F_{\nu}(1400\ \text{\AA})/F_{\nu}(2500\ \text{\AA}),
\end{equation}

\noindent where $F_{\nu}(1400\ \text{\AA})$, $F_{\nu}(2500\ \text{\AA})$ are the continuum flux densities at rest-frame 1400\,\AA\ and 2500\,\AA\, respectively \citep{gibson2008}. Here, we measured the flux densities at  1400 \AA\ and 2500 \AA\ from the best-fit continuum model for J1613+5427, obtaining $R_{\rm{red}} = 0.73$. This value is only $\sim 4$\% higher than the median $R_{\rm{red}}$ of 0.70 reported by \citet{wu2010} for 61 mini-BAL quasars.

Several observational studies have discussed the dependence of strong outflows in NLS1s on the Eddington ratio \citep[e.g.,][]{marziani2003,aoki2005, Komossa2008}. They found a positive correlation between the presence of strong outflow gas and the Eddington ratio in NLS1s. As we discussed above, J1613+5427 has a small BH mass and a high Eddington ratio, even accounting for the wavelength region masked due to absorption. Its mini-BAL outflow may have resulted from radiation pressure due to the high Eddington ratio. Similar trends have been reported for NLS1s in previous studies.

Although the outflow velocity in J1613+5427 is not notably high, the broad absorption features ---particularly at high ionization levels such as N\,\textsc{v}, and C\,\textsc{iv} ---together with the narrow emission lines on the lower continuum, suggest that these NLQ1s are in a highly active evolutionary stage. This stage, which may evolve to typical bright quasars, is characterized by relatively gas-rich environments and high gas covering fractions.

\section{Summary}

This paper presents the initial results of the PFS-SSP survey. The optical data are already useful for detecting a rare population, such as objects with outstanding emission lines. Our primary goal is to search for objects sharing spectral properties with local NLS1s in the intermediate to high-redshift universe, where little is known about NLS1s and their analogs. We searched among 57 broad-line AGNs observed by the PFS-SSP with the requirement of Ly$\alpha$ FWHM $< 2000 \ \mathrm{km\ s^{-1}}$. As a result, we identified four NLQ1s, including a promising candidate, at $z=2.5$ -- $6.0$. All identified NLQ1s have narrow Ly$\alpha$ line widths ($\sim$ 1400 - 1500 $\mathrm{km\ s^{-1}}$) on extremely low continuum. Three objects have strong emission lines with Ly$\alpha$ EW of $\gtrsim 170$\,\AA\  in the rest frame. J1610+5413 and J1610+5401 have outstanding Ly$\alpha$ line luminosities of $\sim 10^{43.7}$ -- $10^{44.2}\ \mathrm{erg\,s^{-1}}$, which corresponds to about $\sim 3$ -- $4$\% of the bolometric luminosities. For three objects, the C\,\textsc{iv} emission line was detected, enabling measurements of the BH masses based on the single-epoch method. They have BH masses of $10^{6.9-7.7}\, M_\odot$, which are close to the values measured in classical NLS1 galaxies in the local universe. We also found that, although we used only Ly$\alpha$ widths for the NLQ1 selection, the C\,\textsc{iv}-based BH masses of the NLQ1s tend to be smaller by $\sim1.5$\, dex than those of other quasars observed by PFS at $2 < z < 6$.

J1613+5427 has BAL features in N\,\textsc{v} and C\, \textsc{iv} with blue continuum, which is also linked to local NLS1 galaxies.

We conclude that NLQ1s may represent the evolutionary phase when a nascent AGN -- still modest in its black hole mass -- begins to awaken, as its maturing accretion disk sweeps away the surrounding gas and the quasar slowly rises to its full power.

\begin{ack}

The instrument ʻ\=Onohiʻula Prime Focus Spectrograph (PFS) including both hardware and software was developed by the PFS collaboration to which over 25 institutes across multiple countries, where the technical activities were conducted by (in alphabetical order) Academia Sinica Institute of Astronomy and Astrophysics (Taiwan), California Institute of Technology, Johns Hopkins University, Kavli Institute for the Physics and Mathematics of the Universe in the University of Tokyo (Kavli IPMU), Laboratoire d'Astrophysique de Marseille, Laboratório Nacional de Astrofísica (Brazil), Max-Planck-Institut für Astrophysik, Max-Planck-Institut für extraterrestrische Physik, NASA Jet Propulsion Laboratory, National Astronomical Observatory of Japan (NAOJ), Princeton University, and Universidade de São Paulo under the oversight by Project Office hosted by Kavli IPMU (later NAOJ). There were also essential commitments from academic and industrial partners such as Durham University (United Kingdom) and Bertin Technologies (France).

The ʻ\=Onohiʻula PFS development work was supported by World Premier International Research Center Initiative (WPI), Ministry of Education, Culture, Sports, Science and Technology (MEXT), Japan. Kavli IPMU was established and supported by World Premier International Research Center Initiative (WPI), MEXT, Japan.
We gratefully acknowledge support from the Funding Program for World-Leading Innovative R\&D on Science and Technology (FIRST) program ”Subaru Measurements of Images and Redshifts (SuMIRe)” by Council for Science and Technology Policy (CSTP), Japan. This work is supported by Japan Society for the Promotion of Science (JSPS) KAKENHI Grant Numbers JP15H05893, JP15K21733, JP15H05892, JP20H05850, JP20H05855, and JP23H05438.
The work at Princeton University, Johns Hopkins University, and California Institute of Technology is supported in part by NSF Award 1636426.
The work in ASIAA, Taiwan, is supported by the Academia Sinica of Taiwan.
The work in Brazil is supported by grants from  CNPq (308994/2021-3) and FAPESP (2011/51680-6).
The work in France is supported by CNRS and Aix Marseille University.

ʻ\=Onohiʻula PFS makes use of the mechanical housing so-called POpt2 that accommodates the Prime Focus Instrument and integrates the Wide Field Corrector lens system generating a flat focal plane with good image qualities across the wide field of view at the Subaru’s prime focus. We appreciate all efforts to make these crucial components of infrastructure operational in conjunction with the development of Hyper Suprime Cam (HSC).
ʻŌnohiʻula PFS software components for instrument control and data processing utilize the platform developed and maintained for Vera C. Rubin Observatory. We appreciate their generosity of making it publicly available as an open source.
We also appreciate the public catalogues from HSC-SSP PDR3, Gaia DR3, and the Pan-STARRS1 Surveys (PS1) which are exploited in PFS observations for field acquisition and auto-guiding of telescope pointing, characterization of sky spectra during exposures, and flux calibrations.

This work is based (in part) on data collected at the Subaru Telescope, which is operated by NAOJ. We are honored and grateful for the opportunity of observing the Universe from Maunakea, which has the cultural, historical, and natural significance in Hawaii.

We appreciate the development and operation of PFS Science Platform by Subaru Telescope and Astronomy Data Center at NAOJ which enables access to both PFS and HSC data and various analyses on the server side.

The HSC collaboration includes the astronomical communities of Japan and Taiwan, and Princeton University. The HSC instrumentation and software were developed by the National Astronomical Observatory of Japan (NAOJ), the Kavli Institute for the Physics and Mathematics of the Universe (Kavli IPMU), the University of Tokyo, the High Energy Accelerator Research Organization (KEK), the Academia Sinica Institute for Astronomy and Astrophysics in Taiwan (ASIAA), and Princeton University. Funding was contributed by the FIRST program from the Japanese Cabinet Office, the Ministry of Education, Culture, Sports, Science and Technology (MEXT), the Japan Society for the Promotion of Science (JSPS), Japan Science and Technology Agency (JST), the Toray Science Foundation, NAOJ, Kavli IPMU, KEK, ASIAA, and Princeton University.

This work has made use of data from the European Space Agency (ESA) mission Gaia (https://www.cosmos.esa.int/gaia), processed by the Gaia Data Processing and Analysis Consortium (DPAC, https://www.cosmos.esa.int/web/gaia/dpac/consortium). Funding for the DPAC has been provided by national institutions, in particular the institutions participating in the Gaia Multilateral Agreement.

The Pan-STARRS1 Surveys (PS1) and the PS1 public science archive have been made possible through contributions by the Institute for Astronomy, the University of Hawaii, the Pan-STARRS Project Office, the Max Planck Society and its participating institutes, the Max Planck Institute for Astronomy, Heidelberg, and the Max Planck Institute for Extraterrestrial Physics, Garching, The Johns Hopkins University, Durham University, the University of Edinburgh, the Queen’s University Belfast, the Harvard-Smithsonian Center for Astrophysics, the Las Cumbres Observatory Global Telescope Network Incorporated, the National Central University of Taiwan, the Space Telescope Science Institute, the National Aeronautics and Space Administration under grant No. NNX08AR22G issued through the Planetary Science Division of the NASA Science Mission Directorate, the National Science Foundation grant No. AST-1238877, the University of Maryland, Eotvos Lorand University (ELTE), the Los Alamos National Laboratory, and the Gordon and Betty Moore Foundation.


KI acknowledges support under the grant PID2022-136828NB-C44 provided by MCIN/AEI/10.13039/501100011033 / FEDER, UE

KGL acknowledges support from JSPS Kakenhi grant Nos. JP18H05868, JP19K14755 and JP24H00241.

ARA acknowledges partial support from Conselho Nacional de Desenvolvimento Científico e Tecnológico (CNPq) through grant 313739/2023-4.

The first author thanks Takaho Masai and Steven Benton for carefully reading the manuscript and providing valuable comments on the English writing.

\end{ack}




\section*{Appendix 1: Six prestacked spectra of J1611+5359}
Figure 6 shows six prestacked spectra of $z = 6$ PFS-NLQ1 candidates. We confirmed that Ly$\alpha$ emission lines were detected in all frames.

\section*{Appendix 2: Rest-UV spectra of the NLQ1s around Ly$\alpha$ with best-fit models.}
Figure 7 shows quality assessment plots of Ly$\alpha$ fitting for four NLQ1s (Section 3.1).

\section*{Appendix 3: Rest-UV spectra of the NLQ1s around C\,\textsc{iv} with best-fit models.}
Figure 8 shows quality assessment plots of C\,\textsc{iv} fitting for three NLQ1s (Section 4.1).

\bibliographystyle{apj}
\bibliography{reference}

\clearpage

\begin{figure*}[!h]
    \begin{minipage}{0.5\textwidth}
        \includegraphics[width=\textwidth]{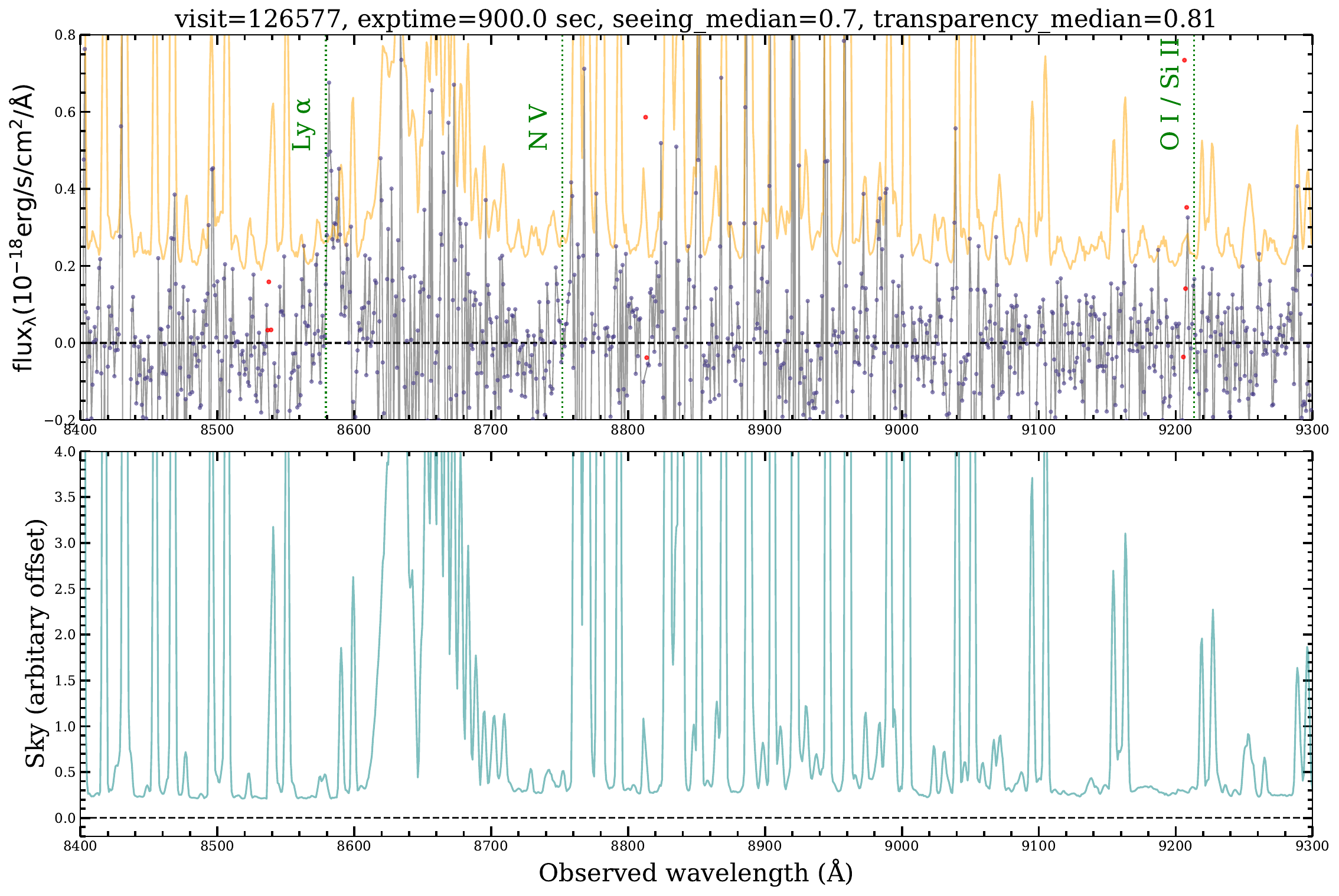}
    \end{minipage}
    \hfill
    \begin{minipage}{0.5\textwidth}
        \includegraphics[width=\textwidth]{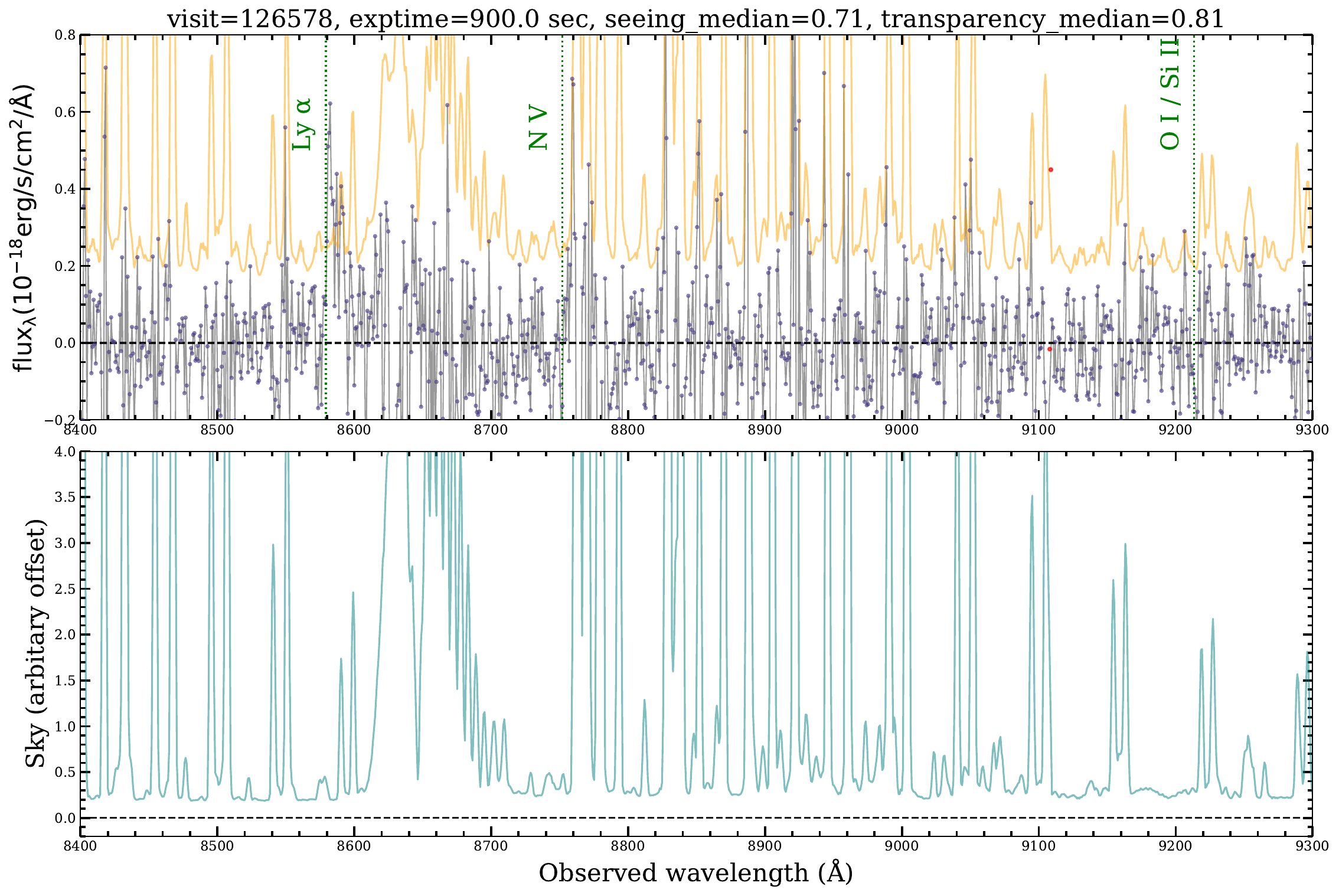}
    \end{minipage}
    \vspace{10pt}

    \begin{minipage}{0.5\textwidth}
        \includegraphics[width=\textwidth]{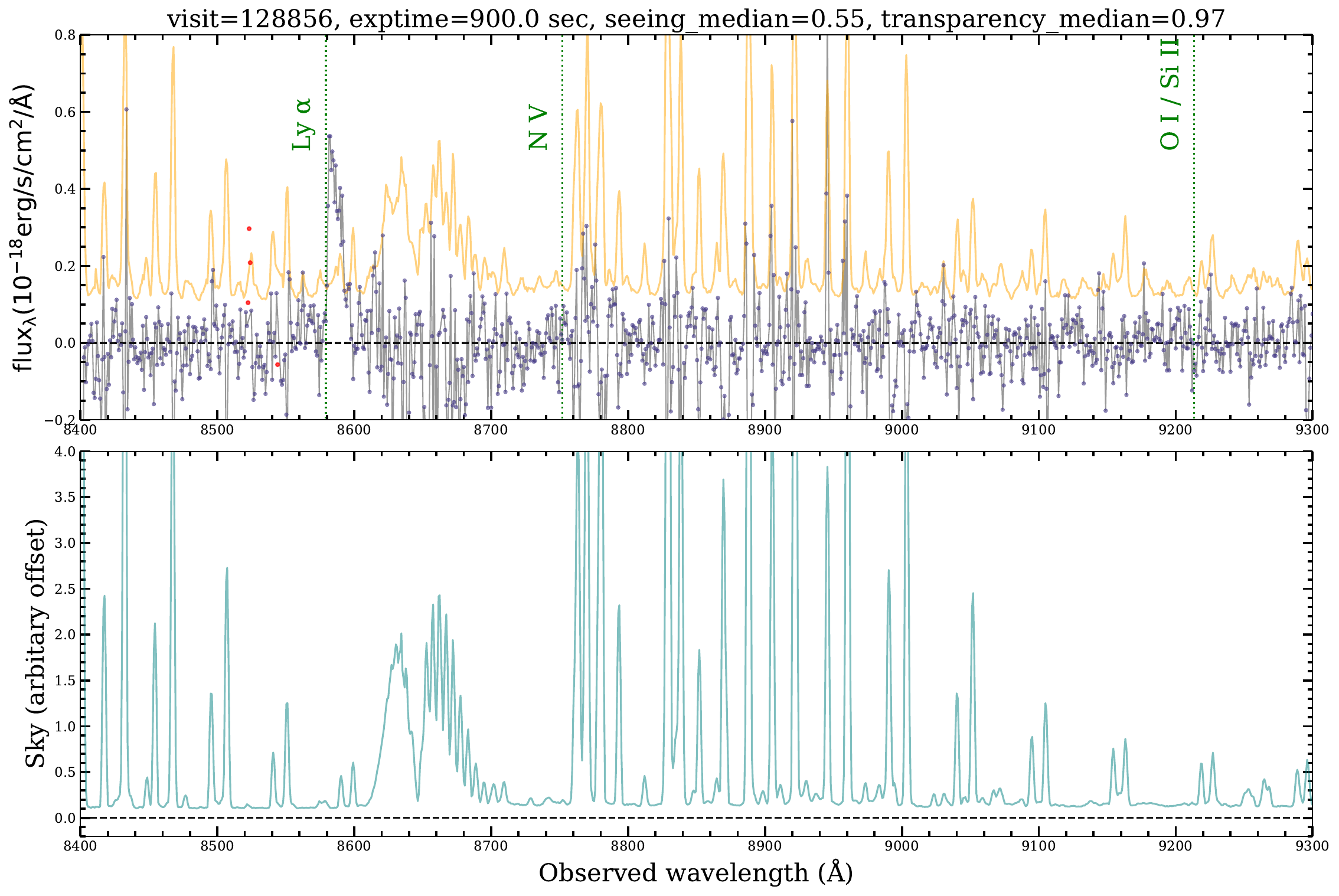}
    \end{minipage}%
    \hfill
    \begin{minipage}{0.5\textwidth}
        \includegraphics[width=\textwidth]{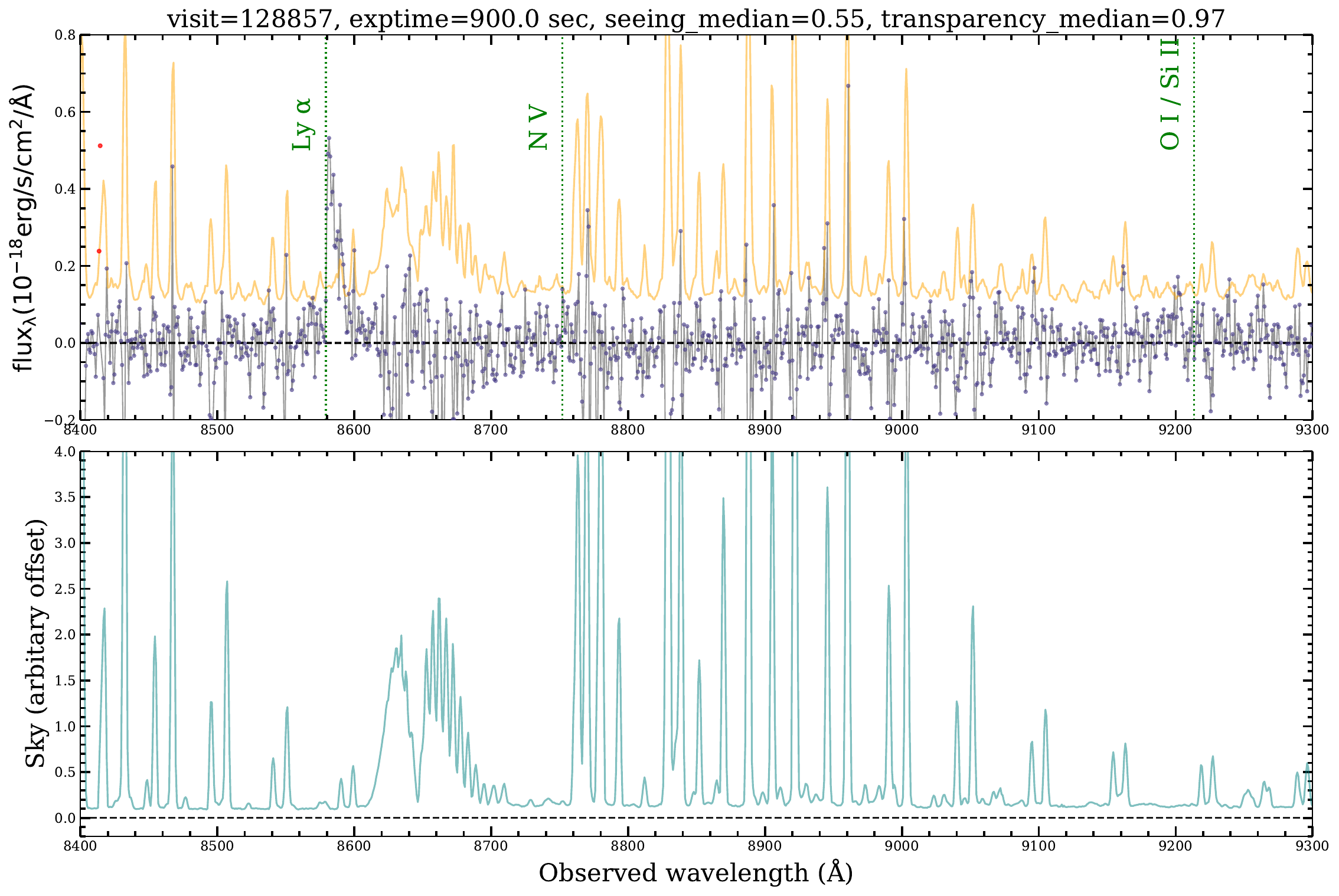}
    \end{minipage}
    \vspace{10pt}

    \begin{minipage}{0.5\textwidth}
        \includegraphics[width=\textwidth]{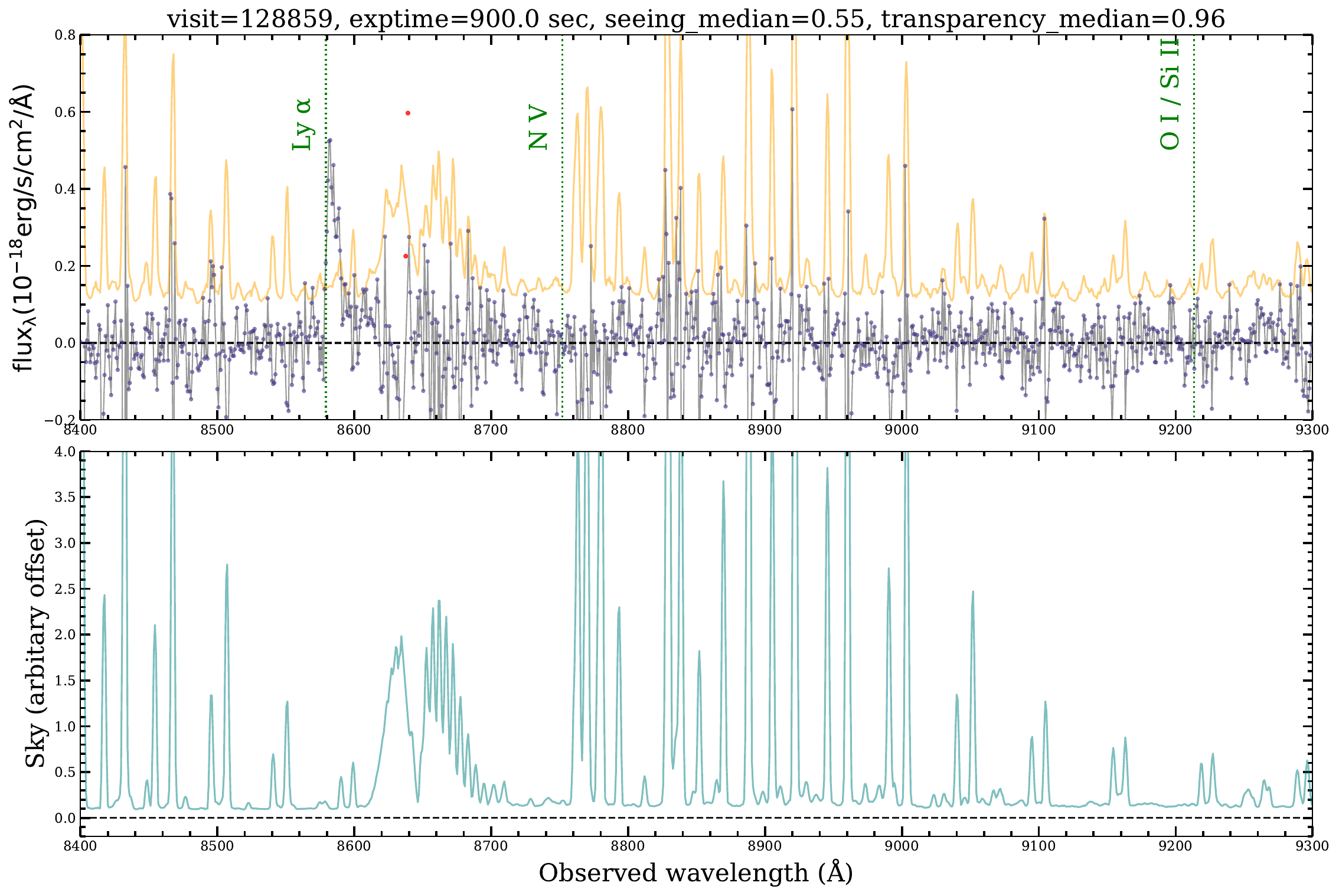}
    \end{minipage}%
    \hfill
    \begin{minipage}{0.5\textwidth}
        \includegraphics[width=\textwidth]{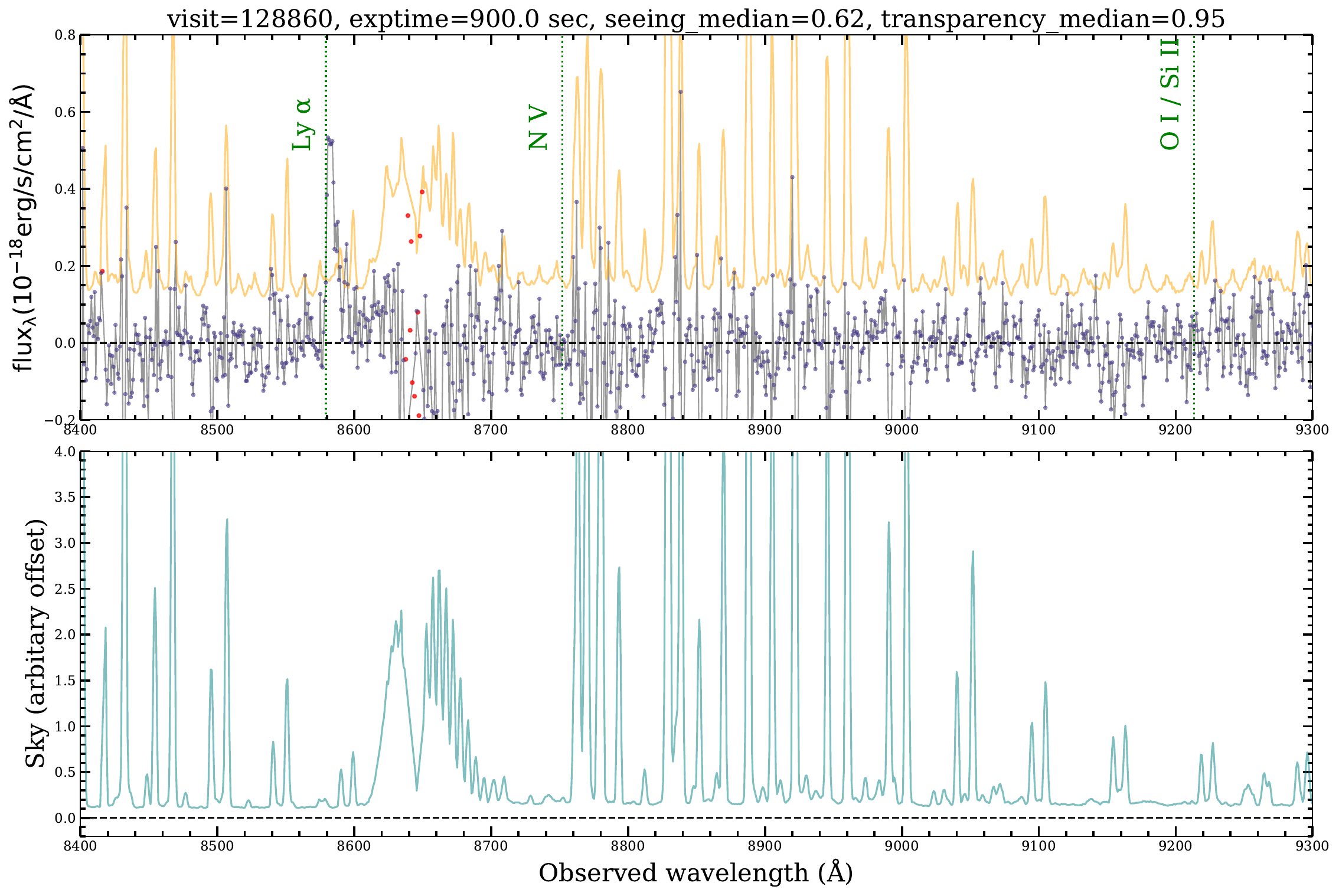}
    \end{minipage}

    \caption{Six prestacked spectra of $z=6$ PFS-NLQ1 candidate (J1611+5359). Observed fluxes are represented by black lines and dark blue dots. Errors are represented by orange lines. Sky spectra are shown in light blue. Bad pixels are shown by red dots. The green dotted lines indicate the expected positions of common emission lines. We also show the exposure (visit) ID, exposure time, and median seeing above each panel.
    {Alt text: A plot with two panels sharing the same horizontal axis of observed wavelength in angstroms, from about 8400 to 9300. The top panel shows the observed spectrum in units of 10 to the power of negative 18 erg per second per square centimeter per angstrom on the vertical axis. The bottom panel shows the sky background spectrum in arbitrary units.}}
    \label{fig:prestack}
\end{figure*}

\begin{figure*}[h]
\centering
\includegraphics[width=15cm,height=15cm]{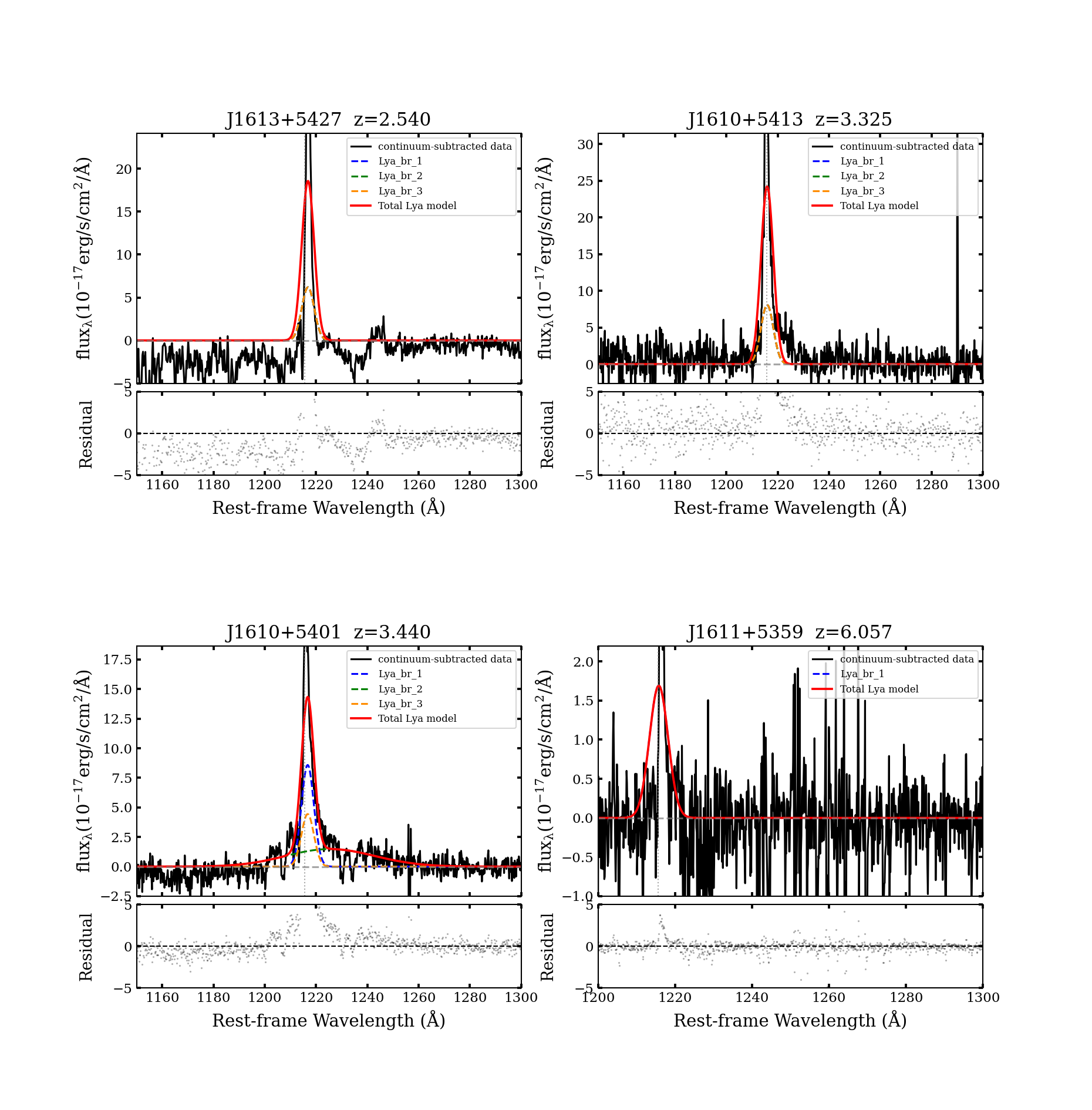}
\caption{Rest-UV spectra of the NLQ1s around Ly$\alpha$. The ordinate is the continuum-subtracted flux density in units of $10^{-17} \rm{erg\  s^{-1} \ cm^{-2}}$\  \AA$^{-1}$. The red lines represent the best-fit models obtained by \texttt{PyQSOFit}, composed of broad components (blue, green, and orange dashed lines). For J1611+5359 ($z$=6.06), the fitting was performed at wavelengths redwards of the observed line peak. The gray dotted lines and the gray dashed lines indicate the expected central wavelengths of Ly$\alpha$, and the zero flux level, respectively. Residuals are also shown at the bottom.
{Alt text: Four line graphs showing the best-fitting result of our model to the data of four narrow-line type-1 quasars around Lyman-alpha.}
}
\label{fig:qsofit2}

\end{figure*}

\begin{figure*}[h]
\centering
\includegraphics[width=0.9\columnwidth]{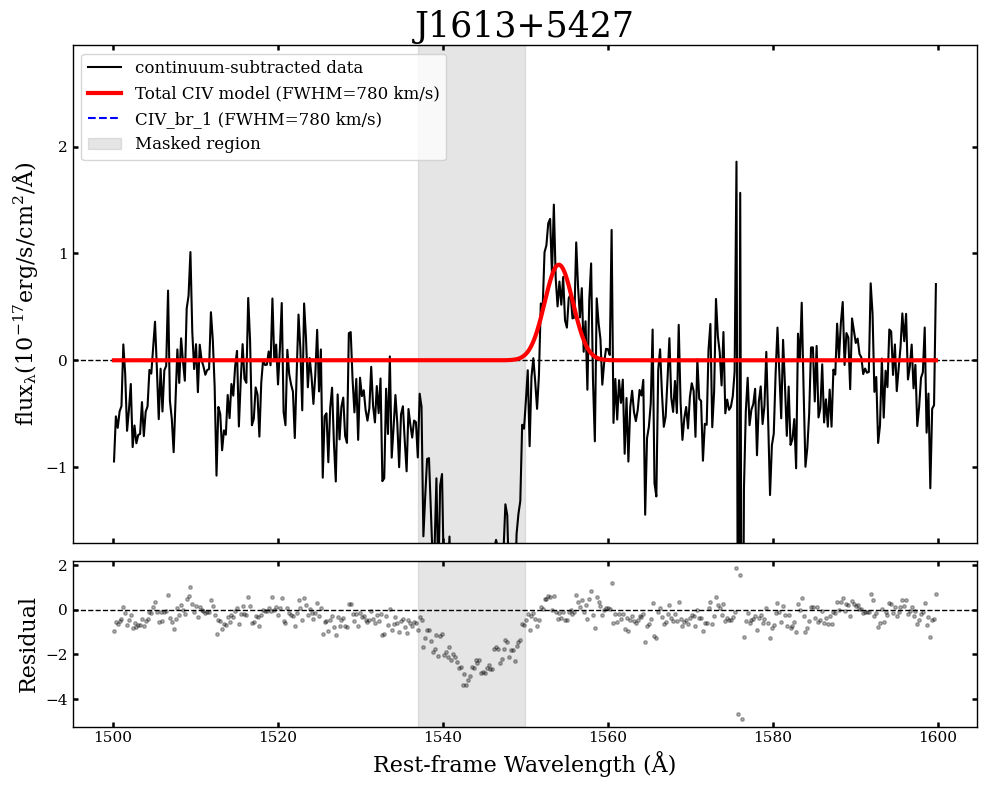}
\includegraphics[width=0.9\columnwidth]{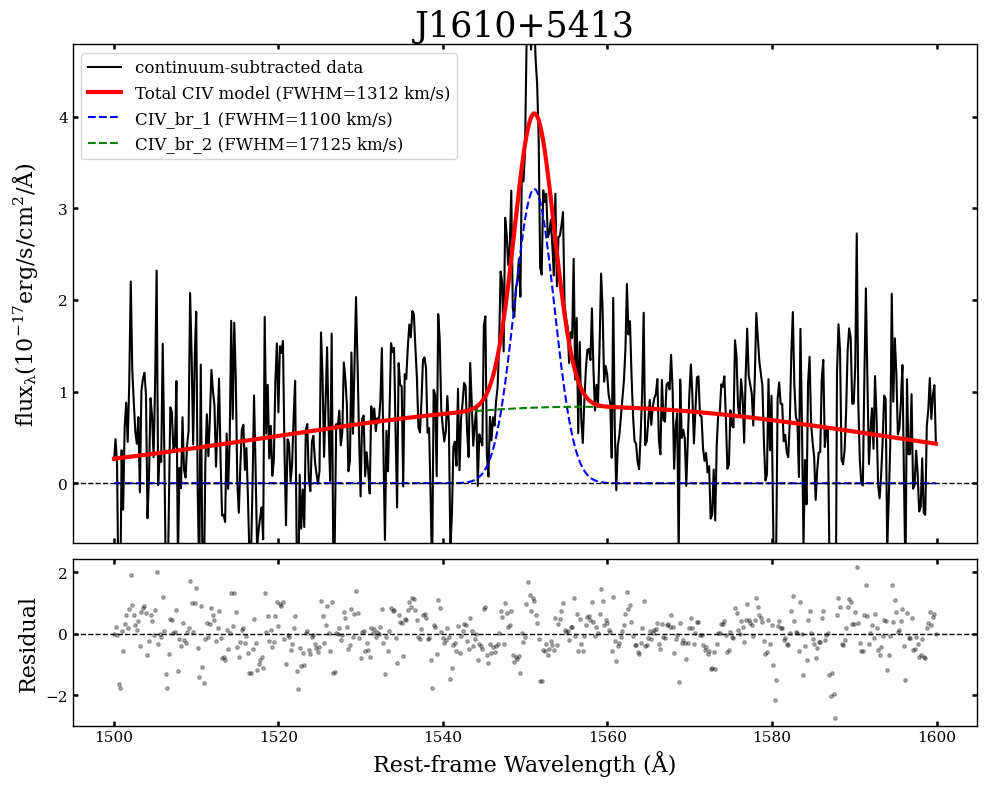}
\includegraphics[width=0.9\columnwidth]{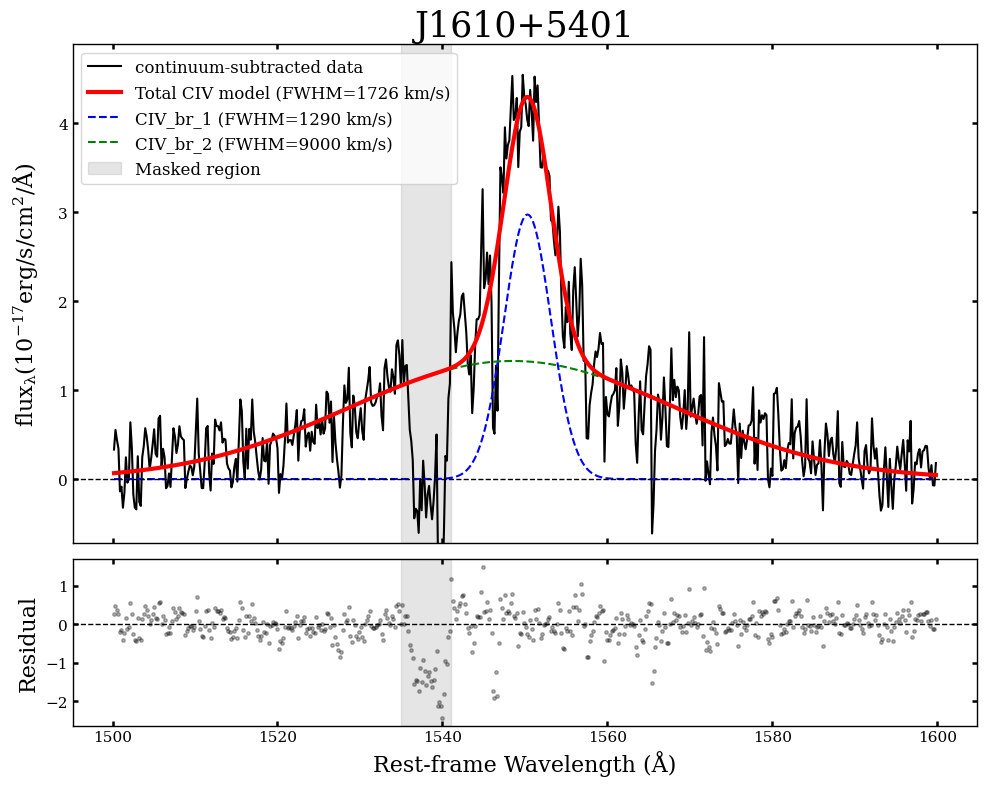}
\caption{Quality assessment plot of C\,\textsc{iv} fitting for J1613+5427 in the upper left, for J1610+5413 in the upper right, and for J1610+5401 in the bottom. The line fitting code is performed with our own code. Observed data in the rest frame are shown with black solid lines. 
The individual Gaussian components are represented by blue and green dashed lines. The best-fit total models are shown with red lines. Gray highlights indicate the masked region in the fitting. Residuals are also shown at the bottom. 
{Alt text: Three line graphs showing the best-fitting result of our model to the data of three narrow-line type-1 quasars around Carbon IV.}}
\label{fig:civ-fit1}
\end{figure*}





\clearpage



\end{document}